\documentclass[12pt, a4paper]{article}

\pdfoutput=1

\usepackage{amsmath}
\usepackage{amsfonts}
\usepackage{amssymb}
\usepackage{mathtools}
\usepackage{bbm}
\usepackage{verbatim}
\usepackage{dsfont}
\usepackage{booktabs}
\usepackage{slashed}

\usepackage{epsfig}
\usepackage{color}
\usepackage[table]{xcolor}
\usepackage{graphicx}
\usepackage{caption}
\usepackage{subcaption}
\usepackage{float}
\usepackage{overpic}
\usepackage{units}

\usepackage{enumerate}
\usepackage{hhline}
\usepackage{multirow}

\usepackage{cite}
\usepackage{xspace}
\usepackage{setspace}

\usepackage[pdftitle={},
  pdfauthor={},
  pdfsubject={},
  bookmarksopen, bookmarksnumbered, bookmarksopenlevel=2, colorlinks=false, linkcolor=blue, citecolor=blue, urlcolor=blue]{hyperref}

\begin{document}

\thispagestyle{empty}

\begin{flushright}
DESY 16-238\\
\end{flushright}
\vskip .8 cm
\begin{center}
	{\Large {\bf Flavour mixings in flux compactifications}}\\[12pt]

\bigskip
\bigskip 
{
{\bf{Wilfried Buchmuller}\footnote{E-mail:
    wilfried.buchmueller@desy.de}} and
{\bf{Julian~Schweizer}\footnote{E-mail: julian.schweizer@desy.de}}
\bigskip}\\[0pt]
\vspace{0.23cm}
{\it Deutsches Elektronen-Synchrotron DESY, 22607 Hamburg, Germany \\ \vspace{0.2cm}
}
\bigskip
\end{center}

\begin{abstract}
\noindent
A multiplicity of quark-lepton families can naturally arise as zero-modes in flux
compactifications. The flavour structure of quark and lepton mass
matrices is then determined by the wave function profiles of the 
zero-modes. We consider a supersymmetric $SO(10)\times U(1)$ model in six
dimensions compactified on the orbifold $T^2/{\mathbb Z_2}$ with Abelian magnetic
flux. A bulk $\mathbf{16}$-plet charged under the $U(1)$
provides the quark-lepton generations whereas two uncharged
$\mathbf{10}$-plets yield two Higgs doublets. Bulk anomaly
cancellation requires the presence of additional 
$\mathbf{16}$- and $\mathbf{10}$-plets. The corresponding zero-modes
form vectorlike split multiplets that are needed to obtain a successful
flavour phenomenology. We analyze the pattern of flavour mixings for
the two heaviest families of the Standard Model and discuss possible generalizations to
three and more generations.
\end{abstract}

\newpage 
\setcounter{page}{2}
\setcounter{footnote}{0}

{\renewcommand{\baselinestretch}{1.5}
\section{Introduction}
\label{sec:introduction}

The explanation of the masses and mixings of quarks and leptons
remains a challenge for theories which go beyond the Standard
Model (SM). Interesting relations between quark and lepton mass matrices are obtained 
in grand unified theories (GUTs) based on the gauge groups
\(SU(4)\times SU(2)\times SU(2)\) \cite{Pati:1974yy}, \(SU(5)\) \cite{Georgi:1974sy},
\(SO(10)\) \cite{Georgi:1974my,Fritzsch:1974nn} and flipped
\(SU(5)\) \cite{Barr:1981qv,Derendinger:1983aj}, and some
understanding of the hierarchies between quark-lepton generations can
be obtained by means of \(U(1)\) flavour symmetries \cite{Froggatt:1978nt}. 

Extending grand unified theories to higher dimensions offers new
possibilities for symmetry breaking and the doublet-triplet splitting
problem. This has been studied in particular in orbifold GUTs where the colour
triplet partners of the Higgs doublets are projected out from the
spectrum of massless states
\cite{Kawamura:2000ev,Hall:2001pg,Hebecker:2001wq,Asaka:2001eh,Hall:2001xr}. 
Orbifold GUTs can also be obtained as intermediate step towards the
embedding of  the Standard Model into string theories \cite{Kobayashi:2004ud,Forste:2004ie,Hebecker:2004ce,Buchmuller:2007qf}. 

Important progress in understanding the flavour structure of the
Standard Model has also been made in the context of the heterotic
string \cite{Raby:2011jt,Nilles:2012cy} as well as in F-theory
\cite{Heckman:2010bq,Lin:2014qga,Cvetic:2015txa}. 
In string theory Yukawa couplings are dynamical quantities whose
values depend on the vacuum structure of the theory. An interesting example are flux
compactifications where the Yukawa couplings can be calculated as
overlap integrals of wave functions that have non-trivial profiles in
the magnetized extra dimensions \cite{Cremades:2004wa}. In a similar
way, Yukawa couplings of magnetized toroidal orbifolds have been analyzed
\cite{Abe:2015yva,Matsumoto:2016okl,Fujimoto:2016zjs,Kobayashi:2016qag}.
The resulting flavour structure depends on the number of pairs of
Higgs doublets. In the simplest cases it appears difficult to obtain
the measured hierarchies of quark and lepton masses
\cite{Matsumoto:2016okl,Fujimoto:2016zjs}.

Magnetic flux leads to a multiplicity of chiral fermion zero-modes
according to the number of flux quanta, which can be used to explain
the number of quark-lepton generations \cite{Witten:1984dg}. Moreover,
flux is an important source of supersymmetry breaking
\cite{Bachas:1995ik}. Starting from a six-dimensional orbifold GUT
model \cite{Asaka:2003iy} with gauge group \(SO(10)\), we have
considered in \cite{Buchmuller:2015jna} possible effects of an additional \(U(1)\) factor.
Abelian magnetic flux can be used to generate a multiplicity of quark-lepton
families from a charged bulk $\mathbf{16}$-plet. Bulk anomaly
cancellation requires additional $\mathbf{10}$-plets and
$\mathbf{16}$-plets that can be uncharged. The orbifold projection
then leads to split multiplets, which allow for the familiar solution
of the doublet-triplet splitting problem in the Higgs sector. Since the
quark-lepton hypermultiplet carries \(U(1)\) charge, the scalar
superpartners of quarks and
leptons acquire large supersymmetry breaking masses of order the GUT
scale, leading to a picture reminiscent of `split supersymmetry'
\cite{ArkaniHamed:2004fb,Giudice:2004tc}. 

In this paper we study the flavour structure of the orbifold GUT model
\cite{Buchmuller:2015jna}. The magnetic flux leads to a non-trivial
profile of the quark and lepton bulk wave functions whereas
the Higgs zero-modes have a constant bulk profile. Contrary to
previously considered flux compactifications, Yukawa couplings
arise from superpotential terms at the orbifold fixed points, i.e.,
from products of quark-lepton wave functions and not as volume integrals
over products of quark-lepton and Higgs wave functions. Moreover,
mass mixings of quarks and leptons with  split multiplets occur. 
This offers new possibilities to obtain a
realistic pattern of quark and lepton mass matrices.

The paper is organized as follows. In Section~2 we recall the needed
features of the symmetry breaking in the GUT model
\cite{Buchmuller:2015jna} and discuss properties of the zero-mode wave
functions. The main part of the paper are Sections~3 and 4. Here the structure
of the quark and lepton Yukawa couplings and mass mixings is discussed,
and a quantitative description is given for masses and mixings of the
two heaviest SM families. Section~5 summarizes some aspects of
supersymmetry breaking and the Higgs sector.
The appendices A and B give some details on
Wilson line breaking and the complex flavour vectors which determine
Yukawa matrices and mass mixings. 

\section{GUT model and symmetry breaking}
\label{sec:model}

In this section we recall the main features of a six-dimensional $SO(10)$ GUT model 
previously discussed in \cite{Buchmuller:2015jna}. In particular, we
discuss the GUT symmetry breaking by means of Wilson lines, list
all the fields relevant for flavour mixing and discuss the properties
of the zero-mode wave functions that determine the Yukawa matrices and
the mixings with split multiplets.

We start from $\mathcal N = 1$ supersymmetry in six dimensions with
$SO(10) \times U(1)$ gauge symmetry, compactified on the orbifold
$T^2/\mathbb Z_2$. In addition to the $\mathbf{45}$-plet of vector
multiplets the model contains six $\mathbf{10}$-plets and four
$\mathbf{16}$-plets. For this set of bulk fields all irreducible and
reducible $SO(10)$ gauge anomalies cancel \cite{Asaka:2003iy,Buchmuller:2015jna}. It is convenient to
group $6d$ vector multiplets into $4d$ vector multiplets $A =
(A_\mu, \lambda)$ and $4d$ chiral multiplets $\Sigma = (A_{5,6},
\lambda')$, and $6d$ hypermultiplets into two chiral multiplets,
$(\phi, \chi)$ and $(\phi', \chi')$ \cite{ArkaniHamed:2001tb}. Note that $(\phi',\chi')$
transform in the the complex conjugate representation compared to
$(\phi, \chi)$. The origin $\zeta_\mathrm{I} =0$ is a fixed point
under reflections, $Ry=-y$, where $y$ denotes the coordinates of the
compact dimensions. Defining fields on the orbifold such that
\begin{equation}
\begin{split}
A(x, -y) &= A(x, y)\,,  \quad \Sigma(x, -y) = -\Sigma(x, y)\,, \\
\phi(x, -y) &= \phi(x, y)\,,  \quad \phi'(x, -y) = -\phi'(x,y)\,,
\end{split}
\end{equation}
breaks $6d$ $\mathcal N = 1$ supersymmetry to 4d $\mathcal N = 1$
supersymmetry at the fixed point $\zeta_\mathrm{I} =0$.

The bulk $SO(10)$ symmetry can be broken to the Standard Model group
by means of two Wilson lines\footnote{In Refs. \cite{Asaka:2001eh} the breaking of
$SO(10)$ was obtained by considering the orbifold $T^2/{\mathbb
  Z_2}^3$. This is equivalent to the symmetry breaking on $T^2/{\mathbb
  Z_2}$ with two Wilson lines that is considered in this section.}. The fixed points $\zeta_i$, $i =
\mathrm{PS},\mathrm{GG},\mathrm{fl}$  are invariant
under combined lattice translations and reflection: $\hat{T}_i \zeta_i
= \zeta_i$, with $\hat{T}_i = T_i \circ R$, $T_i y = y + \lambda_i$,
where $\lambda_i$ denotes a lattice vector (see
Appendix A). Demanding that fields on the orbifold satisfy the relations
\begin{equation}
\begin{split} \label{vectoreta}
 P_\mathrm{PS} A(x, \hat{T}_\mathrm{PS} y) P_{\mathrm{PS}}^{-1} &=
 \eta_\mathrm{PS} A(x, y)\,,\\
P_\mathrm{GG} A(x, \hat{T}_\mathrm{GG} y) P_{\mathrm{GG}}^{-1} &=
 \eta_\mathrm{GG} A(x, y)\,,
\end{split}
\end{equation}
with matrices $P_\mathrm{PS}$, $P_\mathrm{GG}$ given in Appendix A and
parities $\eta_\mathrm{PS}, \eta_\mathrm{GG} = \pm$,
the gauge group $SO(10)$ is broken to the Pati-Salam
subgroup $G_\mathrm{PS} = SU(4)\times SU(2)\times SU(2)$ and the
Georgi-Glashow subgroup $G_\mathrm{GG} = SU(5)\times U(1)_X$ at the
fixed points $\zeta_\mathrm{PS}$ and $\zeta_\mathrm{GG}$, respectively
(see Fig.~1).
The surviving SM gauge group is obtained as intersection of the
Pati-Salam and Georgi-Glashow subgroups of $SO(10)$,
\begin{align}
G_\mathrm{SM'} = G_\mathrm{PS} \cap G_\mathrm{GG}  
= SU(3)\times SU(2)\times U(1)_\mathrm{Y} \times U(1)_\mathrm{X}\,.
\end{align}
\begin{figure}
\centering 
\begin{overpic}[scale = 0.3, tics=10]{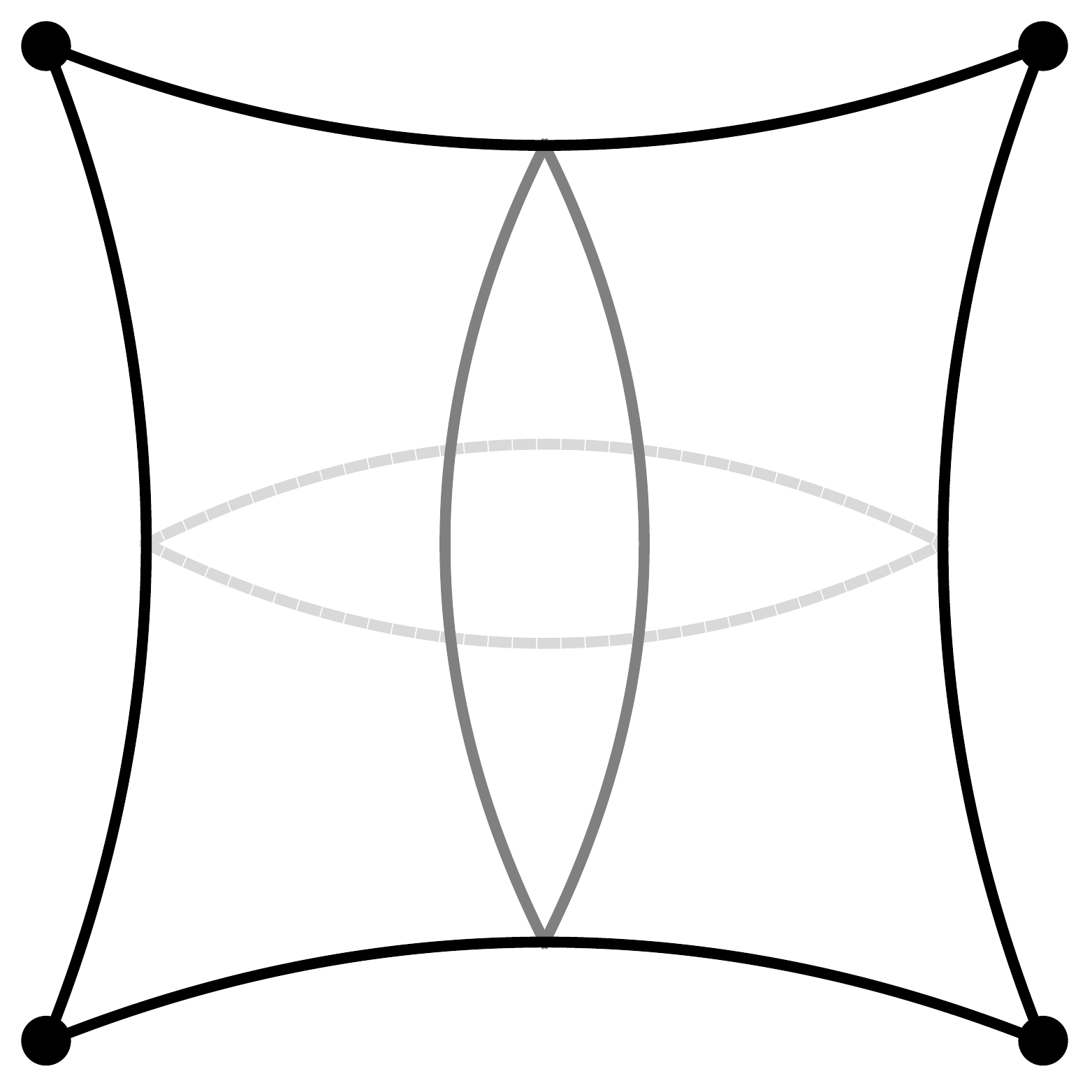}
	\put(-7, 2){$\zeta_{\mathrm{I}}$}
	\put(100, 2){$\zeta_{\mathrm{PS}}$}
	\put(100, 97){$\zeta_{\mathrm{fl}}$}
	\put(-12, 97){$\zeta_{\mathrm{GG}}$}
\end{overpic}
\caption{Orbifold $T^2/{\mathbb Z_2}$  with two Wilson lines and the
fixed points $\zeta_I$, $\zeta_{PS}$, $\zeta_{GG}$, and $\zeta_{fl}$.}
\end{figure}
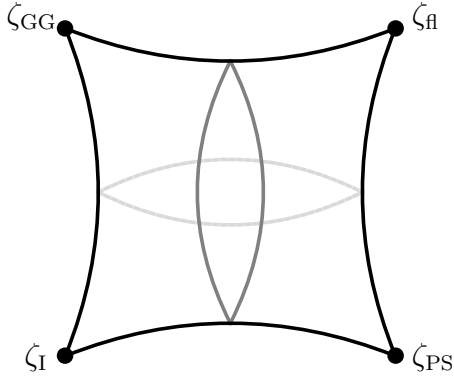
\renewcommand{\arraystretch}{1.2}
\begin{table}
\begin{center}
   $\begin{array}[h]{|c||cc|cc|cc|cc|}\hline
 \mbox{SO(10)} &
     \multicolumn{8}{|c|}{ \mathbf{10} }
     \\ \hline
     G_\mathrm{PS} &
     \multicolumn{2}{|c|}{ ( \mathbf{1}, \mathbf{2}, \mathbf{2}) } &
     \multicolumn{2}{|c|}{ ( \mathbf{1}, \mathbf{2}, \mathbf{2}) } &
     \multicolumn{2}{|c|}{ ( \mathbf{6}, \mathbf{1}, \mathbf{1}) } &
     \multicolumn{2}{|c|}{ ( \mathbf{6}, \mathbf{1}, \mathbf{1}) }
     \\ \hline
     G_\mathrm{GG} &
     \multicolumn{2}{|c|}{ \mathbf{5}^\ast{}_{-2} } &
     \multicolumn{2}{|c|}{ \mathbf{5}{}_{+2} } &
     \multicolumn{2}{|c|}{ \mathbf{5}^\ast{}_{-2} }  &
     \multicolumn{2}{|c|}{ \mathbf{5}{}_{+2} }
     \\ \hline
 {} &
    \eta_\mathrm{PS} & \eta_\mathrm{GG} &
    \eta_\mathrm{PS} & \eta_\mathrm{GG} &
    \eta_\mathrm{PS} & \eta_\mathrm{GG} &
    \eta_\mathrm{PS} & \eta_\mathrm{GG}
    \\ \hline 
     H_1 &
     + & - &
     + & + &
     - & - &
     - & +
     \\ 
   &  \multicolumn{2}{|c|}{} & \multicolumn{2}{|c|}{H_u} &
     \multicolumn{2}{|c|}{} & \multicolumn{2}{|c|}{}
     \\ \hline
     H_2 &
     + & + &
     + & - &
     - & + &
     - & -
     \\ 
   &  \multicolumn{2}{|c|}{H_d} & \multicolumn{2}{|c|}{} &
     \multicolumn{2}{|c|}{} & \multicolumn{2}{|c|}{}
     \\ \hline
     H_3 &
     - & - &
     - & + &
     + & - &
     + & +
     \\ 
   &  \multicolumn{2}{|c|}{} & \multicolumn{2}{|c|}{} &
     \multicolumn{2}{|c|}{} & \multicolumn{2}{|c|}{d}
     \\ \hline
     H_4 &
     - & - &
     - & + &
     + & + &
     + & -
     \\ 
   &  \multicolumn{2}{|c|}{} & \multicolumn{2}{|c|}{} &
     \multicolumn{2}{|c|}{d^c} & \multicolumn{2}{|c|}{}
     \\ \hline\hline
    \mbox{SO(10)} & \multicolumn{8}{|c|}{ \mathbf{16} }
    \\ \hline
    G_\mathrm{PS} &
    \multicolumn{2}{|c|}{ (\mathbf{4}, \mathbf{2}, \mathbf{1}) } &
    \multicolumn{2}{|c|}{ (\mathbf{4}, \mathbf{2}, \mathbf{1}) } &
    \multicolumn{2}{|c|}{ (\mathbf{4}^\ast, \mathbf{1}, \mathbf{2}) } &
    \multicolumn{2}{|c|}{ (\mathbf{4}^\ast, \mathbf{1}, \mathbf{2}) }
    \\ \hline
    G_\mathrm{GG} &
    \multicolumn{2}{|c|}{ \mathbf{10}_{-1} } &
    \multicolumn{2}{|c|}{ \mathbf{5}^\ast{}_{+3} } &
    \multicolumn{2}{|c|}{ \mathbf{10}_{-1 }} &
    \multicolumn{2}{|c|}{ \mathbf{5}^\ast{}_{ +3 }, \mathbf{1}_{ -5 } }
    \\ \hline
 {} &
    \eta_\mathrm{PS} & \eta_\mathrm{GG} &
    \eta_\mathrm{PS} & \eta_\mathrm{GG} &
    \eta_\mathrm{PS} & \eta_\mathrm{GG} &
    \eta_\mathrm{PS} & \eta_\mathrm{GG}
    \\ \hline     
\psi &
     - & + &
     - & - &
     + & + &
     + & -
     \\ 
{} &
    \multicolumn{2}{|c|}{q_i}      &
    \multicolumn{2}{|c|}{l_i}       &
    \multicolumn{2}{|c|}{u_i^c, e_i^c}  &
    \multicolumn{2}{|c|}{d_i^c, n_i^c}
    \\
{} &
    \multicolumn{2}{|c|}{ }      &
    \multicolumn{2}{|c|}{}       &
    \multicolumn{2}{|c|}{u^c, e^c}  &
    \multicolumn{2}{|c|}{}
    \\
\hline
     \Psi &
     - & - &
     - & + &
     + & - &
     + & +
     \\ 
{} &
    \multicolumn{2}{|c|}{ }      &
    \multicolumn{2}{|c|}{}       &
    \multicolumn{2}{|c|}{}  &
    \multicolumn{2}{|c|}{D^c,N^c}
    \\\hline\hline
  \mbox{SO(10)} & \multicolumn{8}{|c|}{ \mathbf{16}^\ast }
    \\ \hline
    G_\mathrm{PS} &
    \multicolumn{2}{|c|}{ (\mathbf{4}^\ast, \mathbf{2}, \mathbf{1}) } &
    \multicolumn{2}{|c|}{ (\mathbf{4}^\ast, \mathbf{2}, \mathbf{1}) } &
    \multicolumn{2}{|c|}{ (\mathbf{4}, \mathbf{1}, \mathbf{2}) } &
    \multicolumn{2}{|c|}{ (\mathbf{4}, \mathbf{1}, \mathbf{2}) }
    \\ \hline
    G_\mathrm{GG} &
    \multicolumn{2}{|c|}{ \mathbf{10}^\ast_{+1} } &
    \multicolumn{2}{|c|}{ \mathbf{5}_{-3} } &
    \multicolumn{2}{|c|}{ \mathbf{10}^\ast_{+1 }} &
    \multicolumn{2}{|c|}{ \mathbf{5}_{ -3 }, \mathbf{1}_{ +5 } }
    \\ \hline
 {\mathrm{parities}} &
    \eta_\mathrm{PS} & \eta_\mathrm{GG} &
    \eta_\mathrm{PS} & \eta_\mathrm{GG} &
    \eta_\mathrm{PS} & \eta_\mathrm{GG} &
    \eta_\mathrm{PS} & \eta_\mathrm{GG}
    \\ \hline 
     \psi^c &
     - & + &
     - & - &
     + & + &
     + & -
     \\ 
{} &
    \multicolumn{2}{|c|}{ }      &
    \multicolumn{2}{|c|}{}       &
    \multicolumn{2}{|c|}{u, e}  &
    \multicolumn{2}{|c|}{}
    \\
\hline
     \Psi^c &
     - & - &
     - & + &
     + & - &
     + & +
     \\ 
{} &
    \multicolumn{2}{|c|}{ }      &
    \multicolumn{2}{|c|}{}       &
    \multicolumn{2}{|c|}{}  &
    \multicolumn{2}{|c|}{D,N}
    \\
\hline
    \end{array}$
    \caption{{\rm PS}- and {\rm GG}-parities for bulk $\mathbf{10}$-plets
      and $\mathbf{16}$-plets.}
    \label{tab:P16}
  \end{center}
\end{table}
\noindent
Group theory implies that $SO(10)$ is broken to flipped $SU(5)$,
$G_\mathrm{fl} = SU(5)'\times U(1)_{X'}$, at $\zeta_\mathrm{fl}$.

Analogously to the vector multiplets the hypermultiplets satisfy the relations
\begin{equation}
\begin{split} \label{hypereta}
 P_\mathrm{PS} \phi(x, \hat{T}_\mathrm{PS} y) &=
 \eta_\mathrm{PS} \phi(x, y)\,,\\
P_\mathrm{GG} \phi(x, \hat{T}_\mathrm{GG} y) &=
 \eta_\mathrm{GG} \phi(x, y)\ ,
\end{split}
\end{equation}
where the matrices $P_\mathrm{PS}$ and $P_\mathrm{GG}$ now depend on the
representation of the hypermultiplet. 
The $SO(10)$ multiplets $A$ and $\phi$ can be decomposed into SM
multiplets, $A = \{A^\alpha\}$ and $\phi = \{\phi^\alpha\}$.
Each of them belongs to a repesentation of
$G_\mathrm{PS}$ as well as $G_\mathrm{GG}$ and is therefore
characterized by two parities,
\begin{align}
 A^\alpha(x, \hat{T}_\mathrm{PS} y) &= \eta^\alpha_\mathrm{PS} A^\alpha(x, y)\,,\quad 
A^\alpha(x, \hat{T}_\mathrm{GG} y) = \eta^\alpha_\mathrm{GG} A^\alpha(x, y)\,,\\
\phi^\beta(x, \hat{T}_\mathrm{PS} y) &= \eta^\beta_\mathrm{PS} \phi^\beta(x, y)\,,\quad
\phi^\beta(x, \hat{T}_\mathrm{GG} y) =\eta^\beta_\mathrm{GG} \phi^\beta(x, y)\,.
\end{align}
The parities of the vector multiplet are fixed by the requirement that
the SM gauge bosons are zero-modes. The parities of the
hypermultiplets can be freely chosen subject to the requirement of
anomaly cancellations. A given set of parities then defines a 4d
model with SM gauge group.

Magnetic flux is generated by a $U(1)$  background gauge field. One
bulk $\mathbf{16}$-plet, $\psi$, carries $U(1)$ charge. The other
$\mathbf{16}$-plets, $\psi^c$, $\Psi$ and $\Psi^c$, and the $\mathbf{10}$-plets
$H_1, \ldots, H_6$ have no $U(1)$ charge. Each hypermultiplet
leads to a `split multiplet' of 4d zero-modes that have both parities
positive. This allows for the wanted
doublet-triplet splitting in the Higgs sector. The parities
$\eta^\beta_\mathrm{PS}$ and $\eta^\beta_\mathrm{GG}$ can be chosen
such that $H_1$ and $H_2$ contain the Higgs doublets $H_u$ and $H_d$,
respectively. The  $\mathbf{16}$-plets $\Psi$ and $\Psi^c$ contain zero-modes
$D^c,N^c$ and $D,N$, respectively. Expectation values of  $N^c$ and $N$ break $U(1)_X$, and therefore
$B-L$. $D^c$ and $D$ have down-quark quantum numbers and acquire mass
by mixing with the zero-modes of the $\mathbf{10}$-plets $H_5$ and
$H_6$ listed in Table~3 of Appendix~A.
The $\mathbf{10}$-plets $H_3$ and $H_4$ also have zero-modes with
down quark quantum numbers. In Table~1 all zero-modes are listed,
which are relevant for our discussion of flavour mixing. 

The charged bulk $\mathbf{16}$-plet $\psi$ yields $N$
$\mathbf{16}$-plets of zero-modes for $N$ flux quanta, independent of
the parity assignements, plus
an additional split multiplet of
zero-modes for which both parities are positive. 
The zero-modes of a charged hypermultiplet have non-trivial wave
function profiles. There are four
possibilies to choose a pair of parities $\eta_\mathrm{PS}$,
$\eta_\mathrm{GG}$. Correspondingly, there are four models that differ
by the parities of the SM components, and therefore the 
assignment of the four types of wave functions to quarks and leptons. 
The four models are listed in Table \ref{tab:parity_assignments}. 

\begin{table}
\centering
\caption{The four different parity assignments for the charged
  $\mathbf{16}$-plet $\psi$, encoded in the parities for the 
$(\mathbf{4},\mathbf{2},\mathbf{1})$-plet at $\zeta_{\mathrm{PS}}$ and the
$\mathbf{10}_{-1}$-plet at $\zeta_{\mathrm{GG}}$, and the resulting wave
function for the different SM fields. The profile of the wave
functions is shown in Fig.~2.}
\label{tab:parity_assignments}
\begin{tabular}{ccc|cccccc}
\toprule
{\rm Model} & $\eta_\mathrm{PS}$  & $\eta_\mathrm{GG}$ & $q$ & $u^c$ & $d^c$ & $l$ & $e^c$ & $n^c$ \\ 
\midrule
I & + & +	& \(\psi_{++}\) & \(\psi_{-+}\) & \(\psi_{--}\) & \(\psi_{+-}\) & \(\psi_{-+}\) & \(\psi_{--}\)\\
II & $-$ & +	& \(\psi_{-+}\)	& \(\psi_{++}\) & \(\psi_{+-}\) & \(\psi_{--}\) & \(\psi_{++}\) & \(\psi_{+-}\)\\
III & + & $-$ & \(\psi_{+-}\) & \(\psi_{--}\)	& \(\psi_{-+}\)	& \(\psi_{++}\)	& \(\psi_{--}\) & \(\psi_{-+}\)\\
IV & $-$ & $-$ & \(\psi_{--}\) & \(\psi_{+-}\) & \(\psi_{++}\) & \(\psi_{-+}\) & \(\psi_{+-}\) & \(\psi_{++}\)\\
\bottomrule
\end{tabular}
\end{table}

For our discussion of flavour mixings we choose model II where $u^c$
and $e^c$ have both parities positive. Hence,  the bulk
field $\psi$ has the decomposition
\begin{equation}
\psi = \sum_{i = 1}^N \left[q_i \psi^{(i)}_{-+} + l_i \psi^{(i)}_{--}
  + \left(d^c_i + n^c_i\right) \psi_{+-}^{(i)} \right] 
+ \sum_{\alpha = 1}^{N+1} \left( u^c_{\alpha} + e^c_{\alpha} \right) \psi^{(\alpha)}_{++}\,.
\end{equation}
\begin{figure}
	\centering
	\begin{tabular}{m{1cm}m{.16 \textwidth}m{.16 \textwidth}m{.16 \textwidth}}
		\toprule
		\(\vert \psi_{++} \vert^2\) &
				\includegraphics[width = .16 \textwidth]{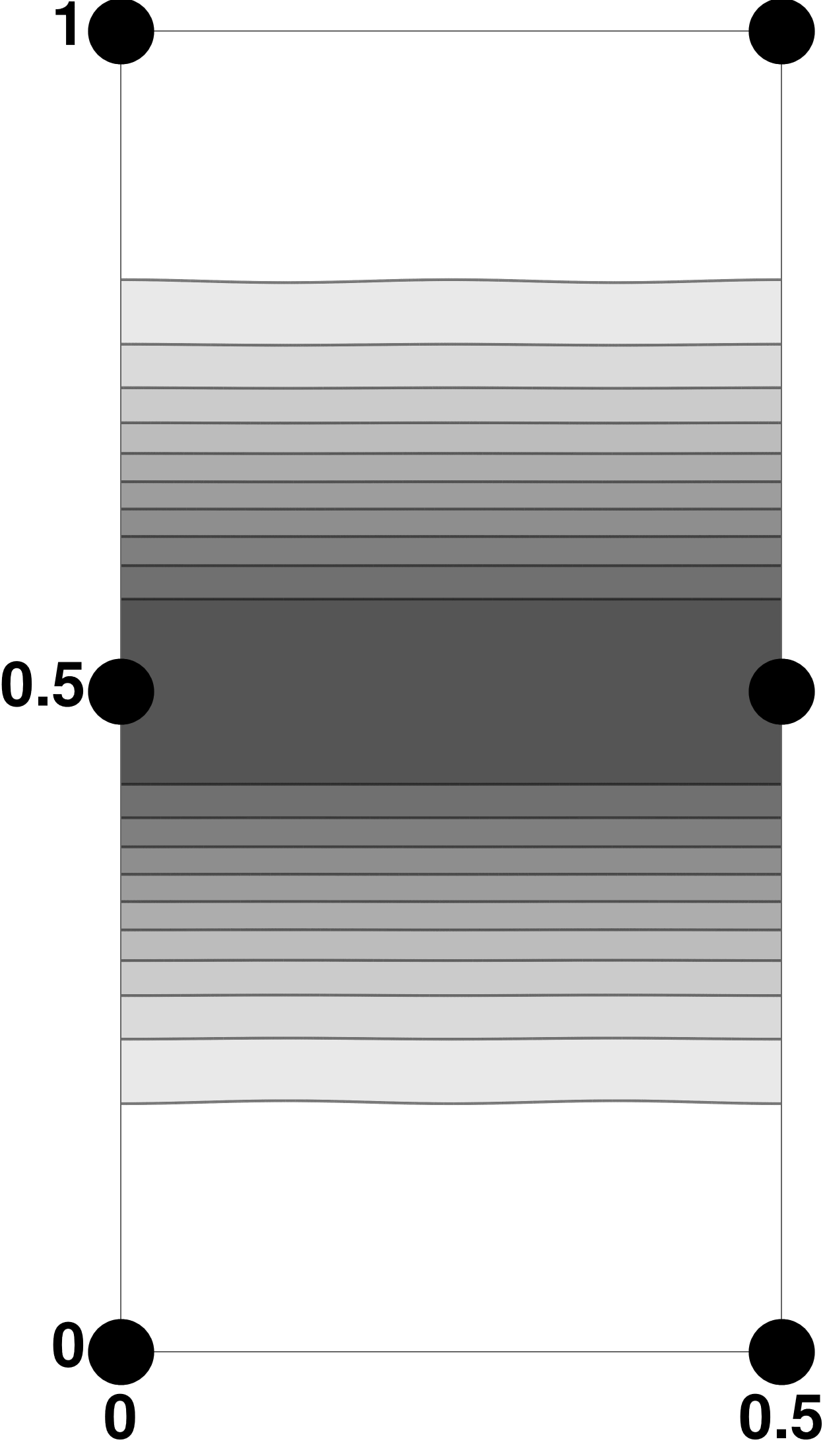} &
				\includegraphics[width = .16 \textwidth]{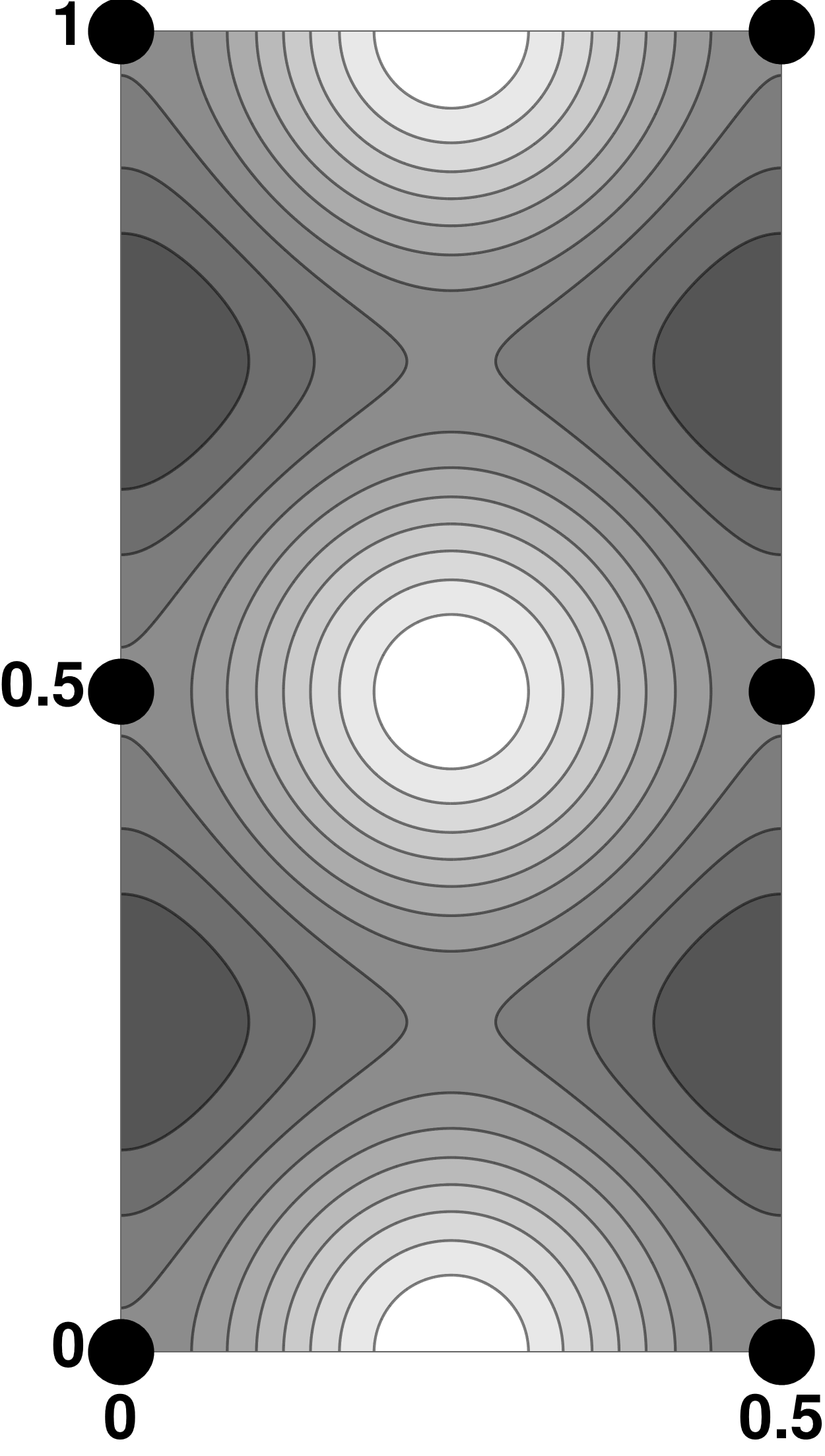} &
				\includegraphics[width = .16 \textwidth]{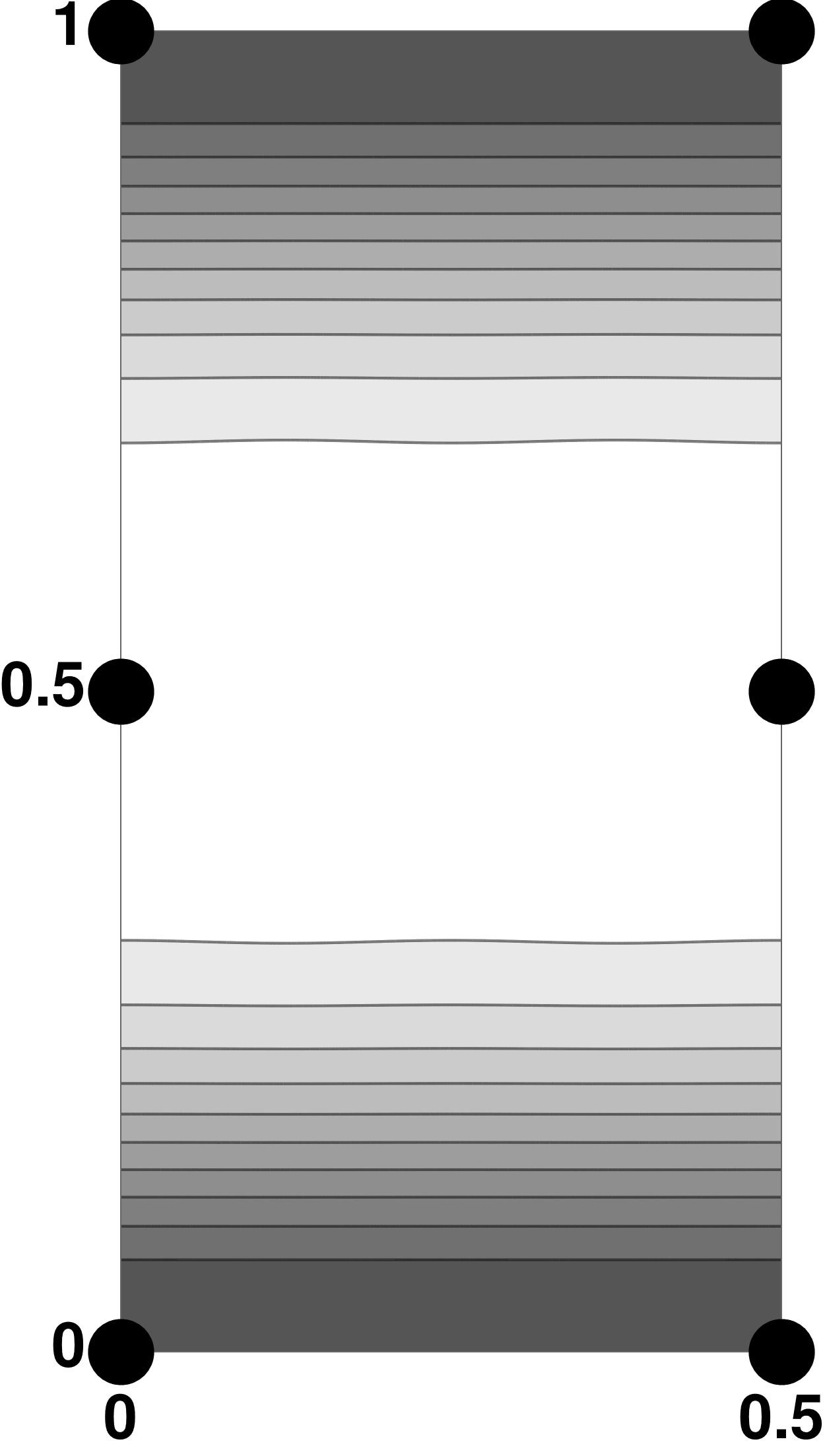}
			 \\\midrule
		\( \vert \psi_{-+} \vert^2 \) &
			\includegraphics[width = .16 \textwidth]{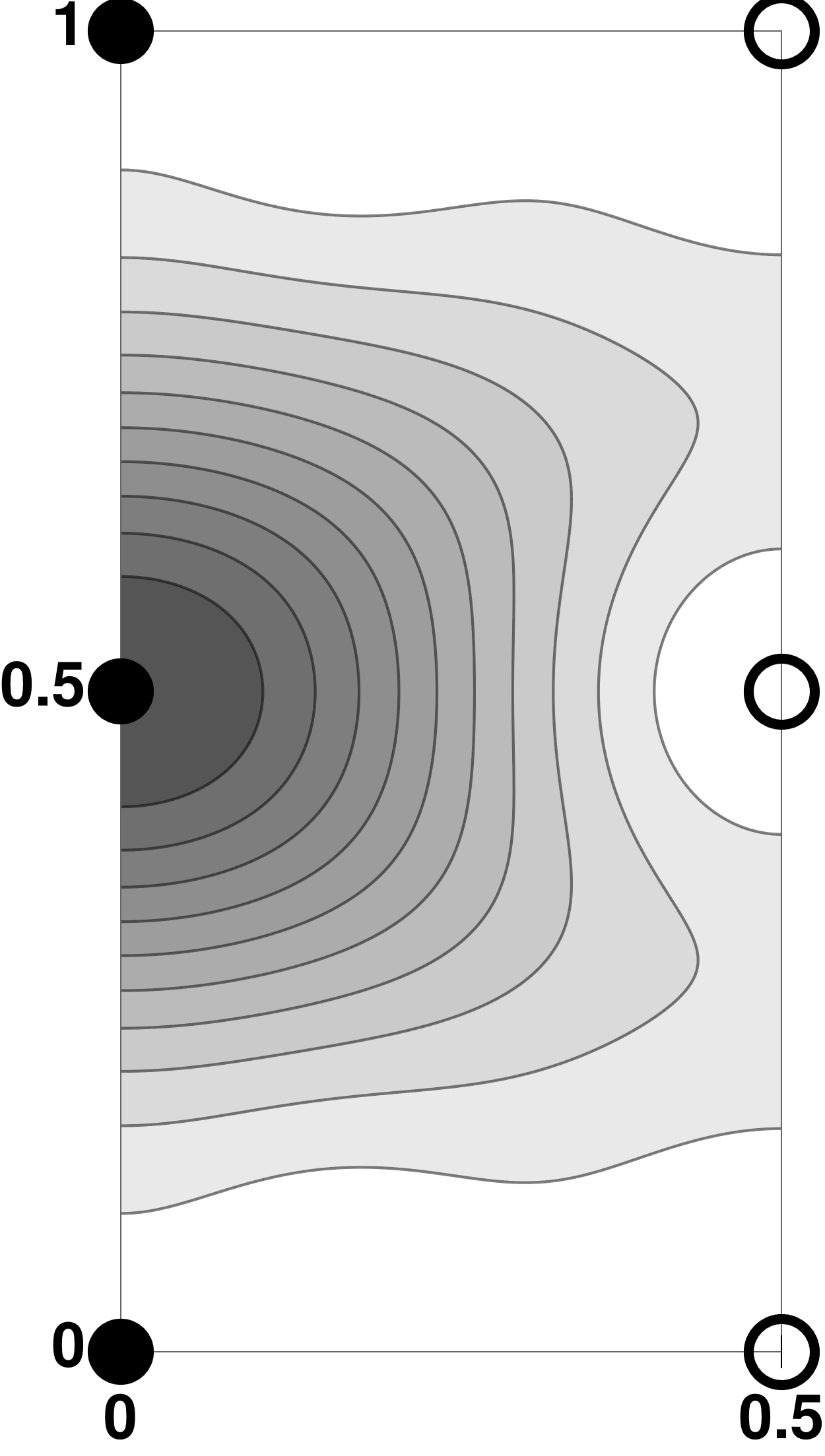} &
			\includegraphics[width = .16 \textwidth]{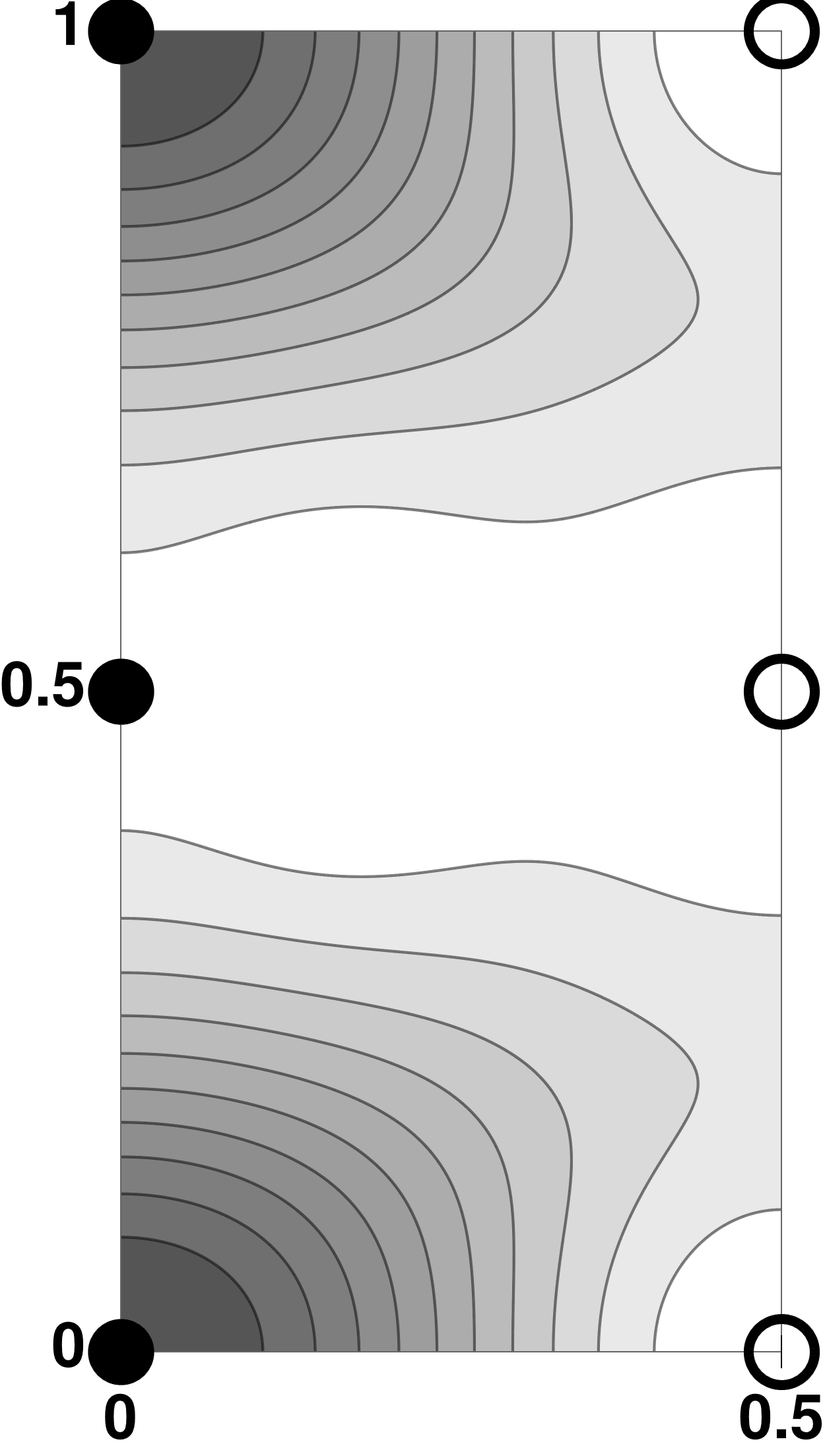}
		\\\midrule
		\( \vert \psi_{+-} \vert^2 \) &
			\includegraphics[width = .16 \textwidth]{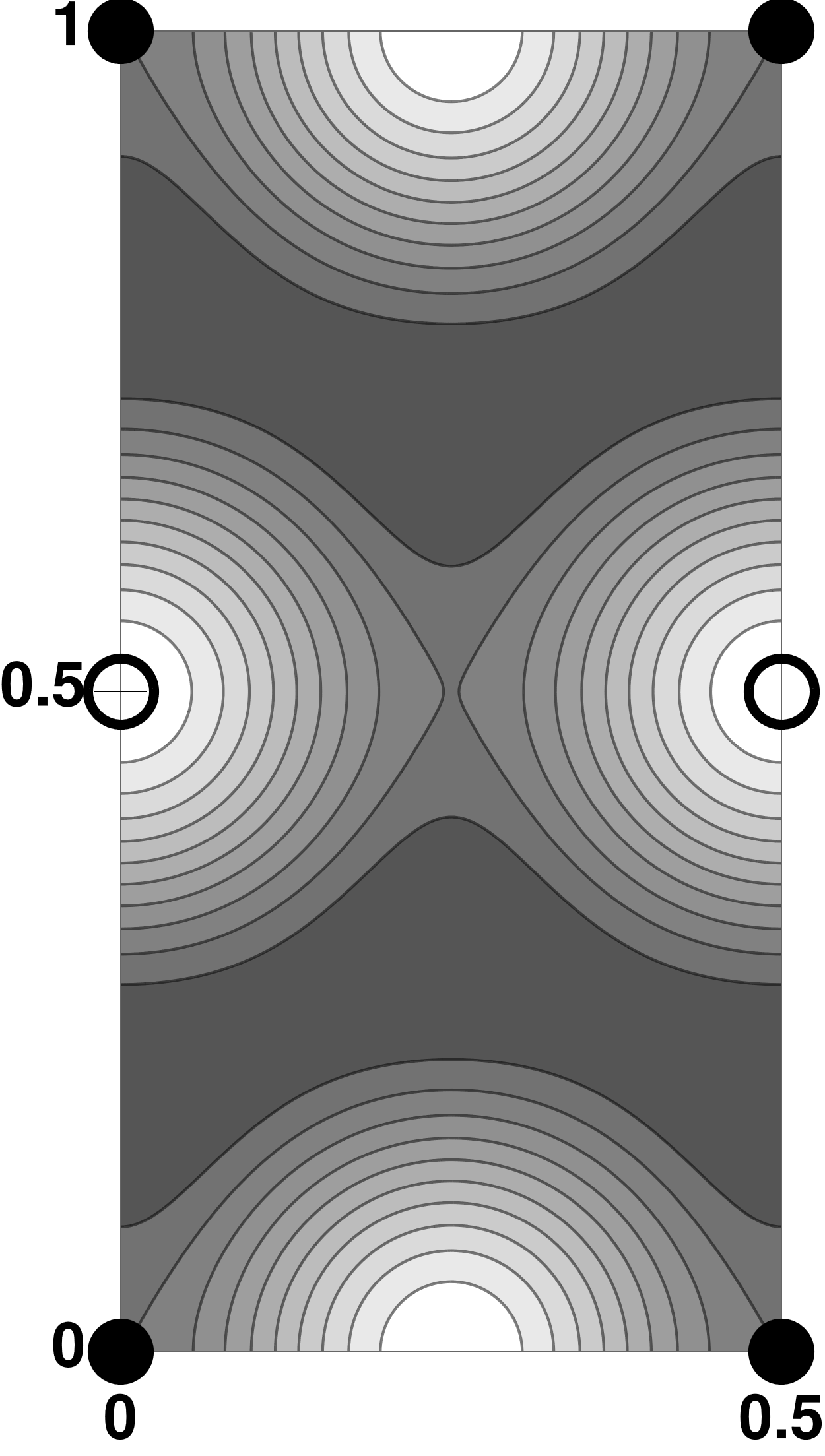} &
			\includegraphics[width = .16 \textwidth]{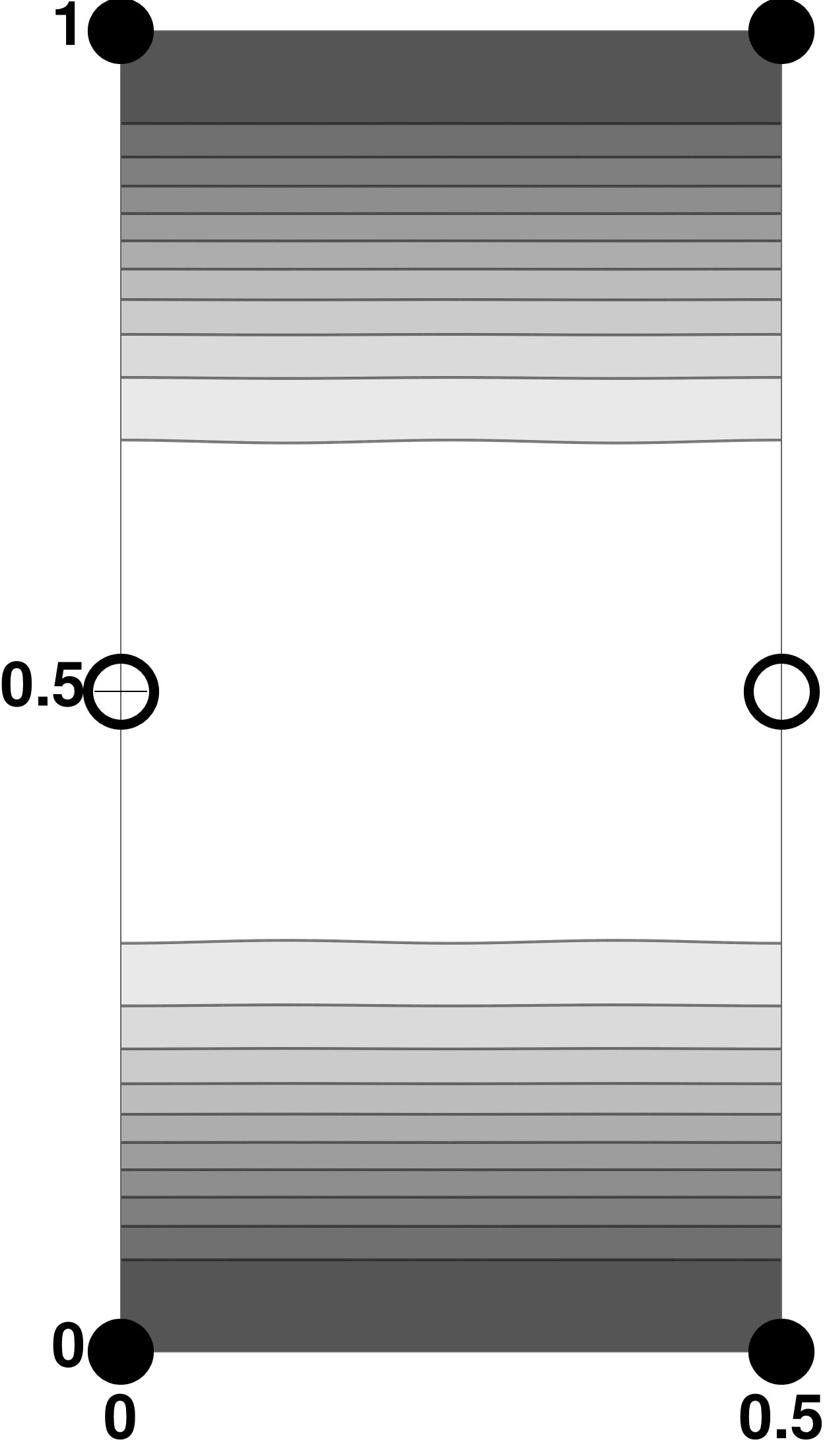}
		\\\midrule
		\(\vert \psi_{--} \vert^2 \) &
			\includegraphics[width = .16 \textwidth]{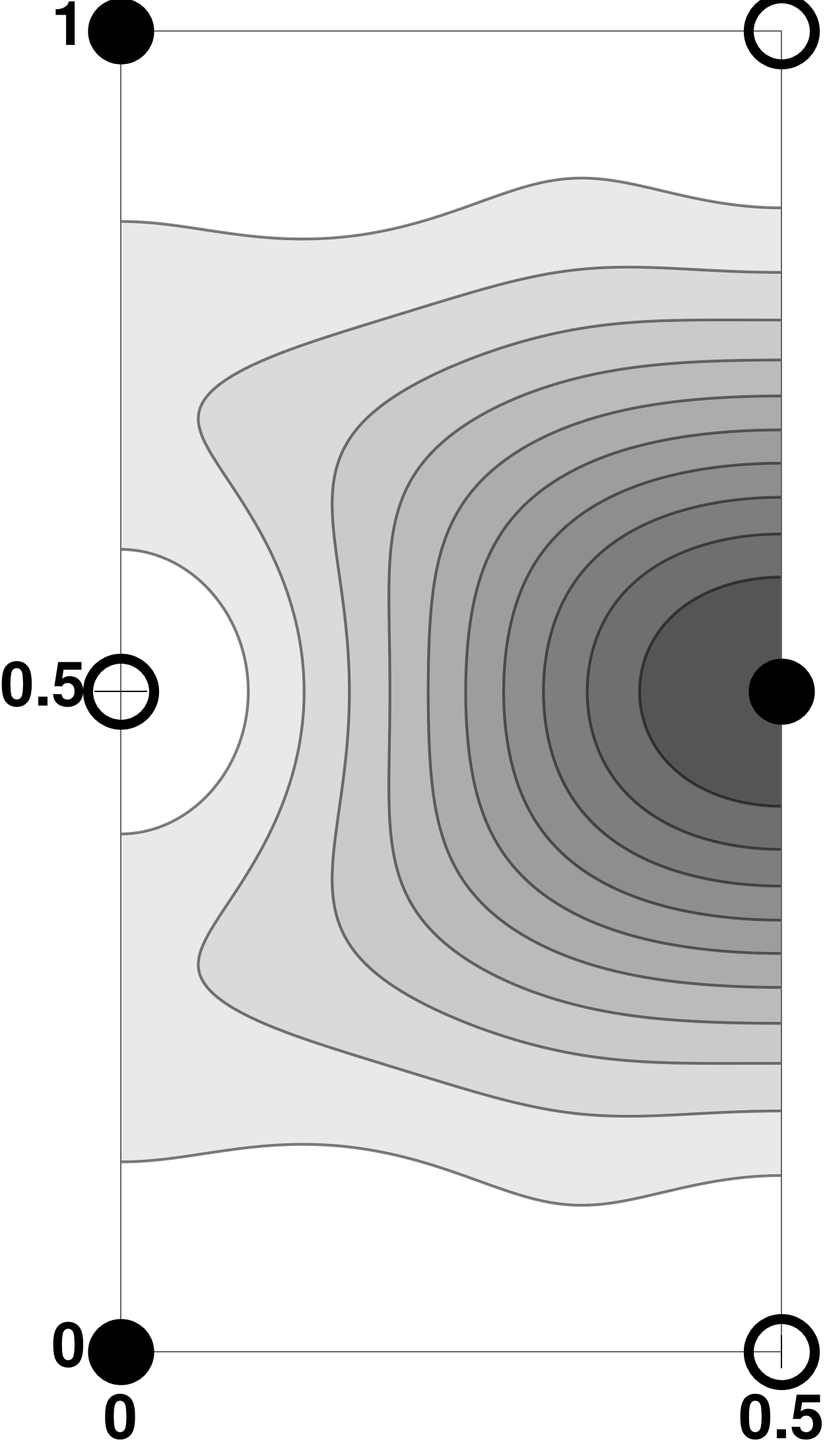} &
			\includegraphics[width = .16 \textwidth]{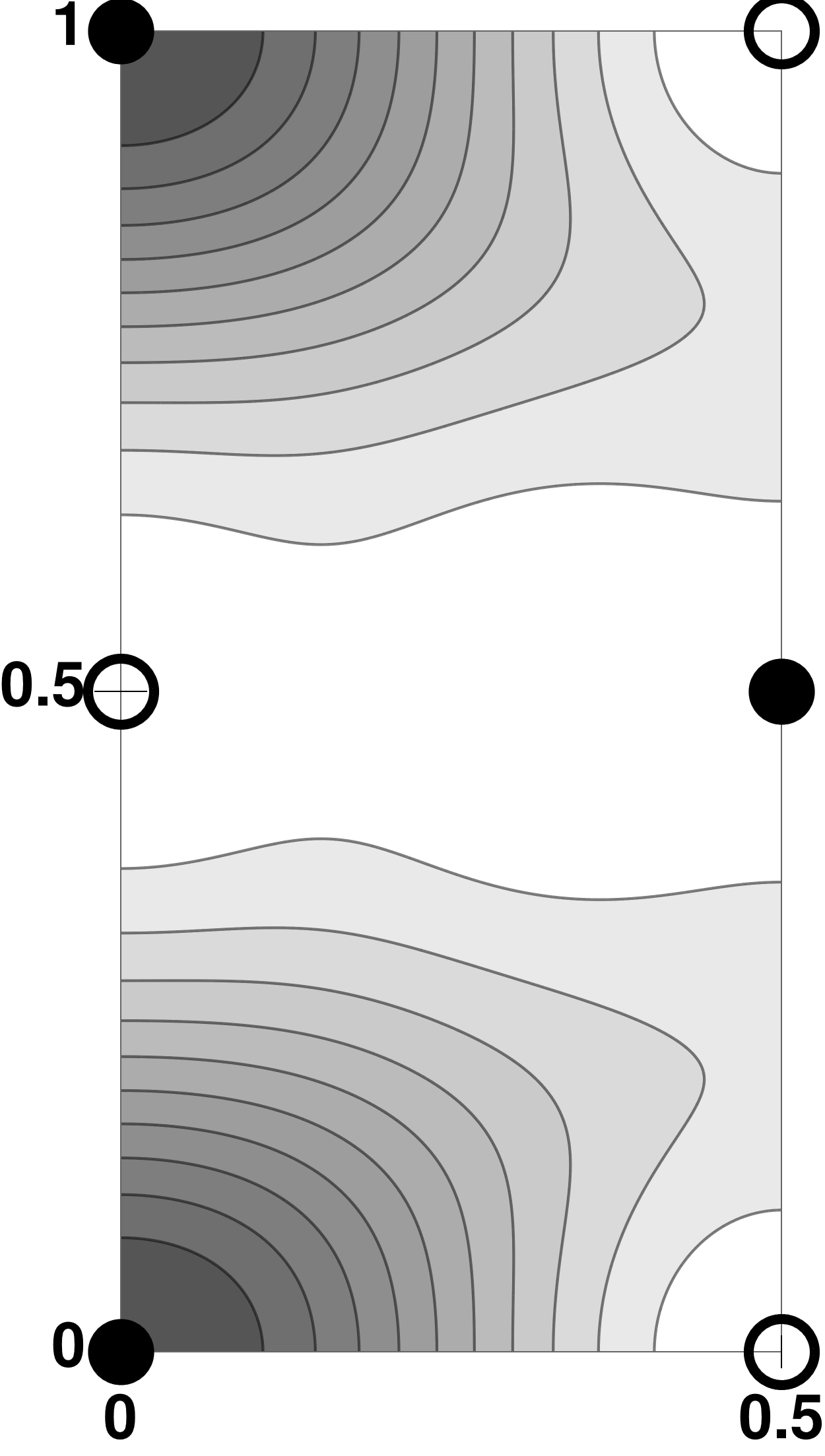}
		\\ \bottomrule
	\end{tabular}
		\caption{The modulus squared of the wave functions
                  \(\psi_{a b}(y)\), $a,b=+,-$, on the
                  orbifold. Darker shades indicate larger values for
                  \(\vert \psi_{a b}\vert^2\). Solid circles represent fixed points with non-vanishing values for the wave function, whereas hollow circles are fixed points where the respective wave function vanishes.}
		\label{fig:wave_functions}
\end{figure}
For the wave functions $\psi^{(j)}_{\eta_\mathrm{PS}
  \eta_\mathrm{GG}}(y)$ we use the expressions given in
\cite{Buchmuller:2015eya}. For $N$ flux quanta they read
\begin{align}
\psi^{(j)}_{\eta_\mathrm{PS} \eta_\mathrm{GG}}(y) = \mathcal N' \, e^{-2\pi  N
  y_2^2 } &\sum_{n \in \mathbb Z} e^{-2 \pi N \left( n - \frac{j}{2N}
  \right)^2 - i \pi \left( n - \frac{j}{2N}\right) (i k_\mathrm{PS} - k_\mathrm{GG})} \nonumber\\
&\times \cos\left[ 2 \pi \left( -2nN + j + \frac{k_\mathrm{PS}}{2}\right) (y_1 + i y_2) \right]\,,\\
\eta_{PS} = &e^{i\pi k_\mathrm{PS}}\,, \quad \eta_\mathrm{GG}
=e^{i\pi k_{GG}}\,, \quad k_\mathrm{PS}, k_\mathrm{GG} = 0, 1\,.\quad \nonumber
\end{align}
For $\eta_\mathrm{PS} = \eta_\mathrm{GG} = +$ there are $N+1$
zero-modes, with $j=0, 1, \ldots N$. In all other cases one has $N$
zero-modes, with $j=0, 1, \ldots N-1$. 
The shape of the wave functions is shown in Fig.~\ref{fig:wave_functions}. The wave functions $\psi^{(j)}_{++}$ are
non-zero at all fixed points. All other wave functions are non-zero at
two fixed points and vanish at the other two. As we shall see in the
subsequent section, this leads to a characteristic pattern of flavour mixings.

\section{Flavour mixings from geometry}
\label{sec:flavor}
In this section we discuss the geometric origin of  flavour mixing
in our model. Supersymmetry in six dimensions
does not allow for a bulk superpotential.
Couplings between the bulk hypermultiplets can therefore only arise at the fixed points.
There, the couplings of fields are proportional to the product of their wave functions, evaluated at the respective location.
The background flux leads to a multiplicity $N$ of fermion families in
our model, and each Yukawa coupling of hypermultiplets at a fixed
point is therefore turned into a $N\times N$ flavour matrix. 
The SM gauge quantum numbers and the Wilson line
configuration determine the wave function for a given field, see Table \ref{tab:parity_assignments}.
In the following we shall focus on model II.

In order to illustrate how the Yukawa matrices arise from the wave
function profiles, let us consider a $2\times 2$ down-type quark Yukawa
matrix. This arises at the \(SO(10)\) fixed point from the term
\begin{equation}\label{down_matrix}
\begin{aligned}
W_I \supset h_d^\mathrm{I} \ \psi \psi\ H_2 \vert_{\zeta_I} &\supset
h_d^\mathrm{I} \  q_i \psi_{-+}^{(i)} \  d^c_j
\psi_{+-}^{(j)}\ H_d\vert_{\zeta_\mathrm{I}} \\
		&\supset v_2 h_d^\mathrm{I}\  d_i d^c_j\
                \psi_{-+}^{(i)} \psi_{+-}^{(j)}
                \vert_{\zeta_\mathrm{I}} 
\equiv  v_2 Y_d^{ij}\ d_i d^c_j \,,
\end{aligned}
\end{equation}
where \(i, j\) are the family indices which label the degeneracy of
of the zero-mode wave functions. 
The wave functions have to be evaluated at the \(SO(10)\) fixed point, and
we assume electroweak symmetry breaking with $v_2 = \langle H_d\rangle$.
From Eq.~\eqref{down_matrix} and Eqs.~\eqref{p-+}, \eqref{p+-}
one obtains the down-quark Yukawa matrix 
\begin{equation}
	\label{eq:down_Yukawa}
Y_d^{ij} = h_d^I \psi_{-+}^{(i)} \psi_{+-}^{(j)}\vert_{\zeta_I} = h_d^I \begin{pmatrix}  
0.32\ (1+ i)  & 0.71 \\
1.50\ (1+ i)  & 3.29  
\end{pmatrix}.
\end{equation}
In principle, there are further contributions to the down-quark Yukawa couplings from other fixed point superpotentials.
However, all operators involving the quark doublet vanish at the
Pati-Salam fixed point, as they are proportional to
\(\psi_{-+}^{(i)}\) that vanishes at \(\zeta_\mathrm{PS}\) due to their negative parity there.
By the same reason there is neither a contribution from the
Georgi-Glashow fixed point where the right-handed down quark wave
functions \(\psi_{+-}^{(i)}\) vanish, nor from the flipped
\(SU(5)'\times U(1)'\) fixed point where both the quark doublets and
the right-handed down quarks cannot couple. Up to higher-dimensional operators,
the model II therefore predicts the down-quark Yukawa couplings
uniquely. 

The matrix given in Eq.~\eqref{eq:down_Yukawa} 
has some characteristic features. First of all, one of the eigenvalues 
vanishes,
\begin{equation}
	\label{eq:YdHierarchy}
	\hat Y_d  = h_d^I \operatorname{diag}(0 ,  5.0)\ ,
\end{equation}
so there is only one heavy state with nonvanishing mass. This follows
immediately  from the fact that the matrix $Y_d$ is a dyadic tensor
build from two vectors that determine the couplings of $q_i$ and $d^c_j$ at the
fixed point $\zeta_I$. As we shall see, a second nonvanishing mass
can be obtained from the mixing with vectorlike split multiplets.
Furthermore, it is interesting that the matrix \eqref{eq:down_Yukawa}
is complex, which is a consequence of the two nonvanishing Wilson
lines.  It therefore naturally incorporates CP violation. 

In most cases, the Yukawa matrices have contributions from two fixed points. 
It is straightforward to list the lowest-dimensional operators that
contribute to fermion masses at the various fixed points after 
\(B-L\) and electroweak symmetry breaking,
\begin{equation}
\label{eq:general_fp}
\begin{aligned}
W_\mathrm{Yuk} = \hphantom{ + } &\delta_\mathrm{I} \big( h_u^\mathrm{I} \, \psi \psi \, H_1 + h_d^\mathrm{I} \, \psi \psi \, H_2 + h_n^\mathrm{I} \, \psi \psi \Psi^c \Psi^c\big) \\
{} + &\delta_\mathrm{PS}  \big( h_u^\mathrm{PS} \, \mathbf 4 \, \mathbf 4^* \, \Delta_1 + h_d^\mathrm{PS}\ \mathbf 4 \, \mathbf 4^* \Delta_2 + h_n^\mathrm{PS}\ \mathbf 4^* \mathbf 4^* F F \big) \\
{} + &\delta_\mathrm{GG} \big( h_u^\mathrm{GG}\, \mathbf{10} \, \mathbf{10} \, H_\mathbf{5} + h_d^\mathrm{GG}\, \mathbf{5}^* \mathbf{10} \, H_{\mathbf 5^*} + h_\nu^\mathrm{GG}\ \mathbf 5^* n^c \, H_{\mathbf{5}} + h_n^\mathrm{GG}\ n^c n^c \, N N \big)\\
{} + &\delta_\mathrm{fl}  \big( h_u^\mathrm{fl} \, \tilde{\mathbf{5}}^* \tilde{\mathbf{10}} \, H_{\tilde{\mathbf{5}}^*} + h_d^\mathrm{fl} \, \tilde{\mathbf{10}} \, \tilde{\mathbf{10}}\, H_{\tilde{\mathbf{5}}} + h_e^\mathrm{fl}\ \tilde{\mathbf{5}}^* e^c\, H_{\tilde{\mathbf{5}}} + h_n^\mathrm{fl} \, \tilde{\mathbf{10}}\, \tilde{\mathbf{10}}\, \tilde T^* \tilde T^* \big)\,,
	\end{aligned}
\end{equation}
where we denote the Higgs fields in the representations of the
unbroken \(SO(10)\) subgroups at the various fixed points by
$H_1\vert_{\zeta_\mathrm{PS}}=\Delta_1$, $H_1\vert_{\zeta_\mathrm{GG}}=H_{\mathbf{5}}$,  
$H_1\vert_{\zeta_\mathrm{fl}}=H_{\tilde{\mathbf{5}}}^*$,  
$H_2\vert_{\zeta_\mathrm{PS}}=\Delta_2$, $H_2\vert_{\zeta_\mathrm{GG}}=H_{\mathbf{5}^*}$,  
$H_2\vert_{\zeta_\mathrm{fl}}=H_{\tilde{\mathbf{5}}}$,  
$\Psi^c\vert_{\zeta_\mathrm{PS}}=F$,
$\Psi^c\vert_{\zeta_\mathrm{GG}}=N$ and
$\Psi^c\vert_{\zeta_\mathrm{fl}}=\tilde{T}^*$;
\(\delta_p = \delta(y-\zeta_p)\), \(p = \mathrm{I}, \ldots , \mathrm{fl}\), and the vacuum expectation values of the
Higgs fields are
 $\langle H_1 \rangle = \langle H_u \rangle = v_1$, $\langle H_2 \rangle
= \langle H_d \rangle =  v_2$ and $\langle \Psi^c \rangle =  \langle N \rangle = v_{B-L}$. 
We refer to the matter fields at the various fixed points by their \(SU(4)\) or \(SU(5)\) representation respectively, marking flipped \(SU(5)'\times U(1)'\) fields with a tilde.
Specifying to model II, all but six of the operators in Eq.~\eqref{eq:general_fp} vanish because in most couplings one of the matter field wave functions is zero at the fixed point. 
This considerably simplifies the superpotential,
\begin{equation}
\label{eq:scenarioII_Yuk}
\begin{aligned}
W_\mathrm{Yuk} =  \hphantom{ + } &\delta_\mathrm{I} \left( h_u^\mathrm{I}\ \psi\psi\ H_1 +
h_d^\mathrm{I}\ \psi\psi\  H_2 + h_n^\mathrm{I}\ \psi\psi \Psi^c\Psi^c\right) \\
+\ &\delta_\mathrm{PS} h_n^\mathrm{PS}\  {\mathbf{4}}^* {\mathbf{4}}^* F F 
+ \delta_\mathrm{GG} h_u^\mathrm{GG}\ \mathbf{10}\ \mathbf{10}\ H_\mathbf{5} 
+ \delta_\mathrm{fl} h_e^\mathrm{fl}\ \tilde{\mathbf{5}}^* e^c\ H_{\tilde{\mathbf{5}}} \,.
\end{aligned}
\end{equation}
The dimension-four operators give  Yukawa couplings for the quarks,
charged leptons and neutrinos, whereas the dimension-five operators
give Majorana masses for the right-handed neutrinos.

As we showed above for the down-type quarks, the wave functions fully determine the matrix structure of the couplings.
Of special note in this regard are the fields with even parities at all fixed points, \(u^c_i\) and \(e^c_i\).
For these, the degeneracy induced by \(N\) flux quanta is \(N+1\)-fold, so the up-quark and charged lepton Yukawa matrices derived from Eq.~\eqref{eq:scenarioII_Yuk} are actually \(N \times (N+1)\) matrices.

As described in the previous section, the considered GUT model
unavoidably predicts additional vectorlike states that are expected
to have mass terms of order the GUT scale. Mixing them with quarks
and leptons, one can obtain realistic mass matrices. 
First, we project the bulk \(\mathbf{16}^*\)-plet, \(\psi^c\), in
such a way that it complements the additional \(u^c\) and \(e^c\)
we obtained from the flux (see Table~1).
Mixing with these zero-modes $u$ and $e$, the up-quark and charged lepton Yukawa
couplings turn into 
\((N+1)\times (N+1)\) matrices.
Next, we project two of the additional bulk \(\mathbf{10}\)-plets to a vectorlike pair of down-type quarks, \(H_3 \supset d\), \(H_4 \supset d^c\).
Introducing mixing in the down-quark sector provides sufficient
freedom to reproduce the measured features of quark mixing.
Finally, a large number of SM singlet fields are required
by the cancellation of gravitational anomalies. 
These can mix with the right-handed neutrinos, and with the left-handed neutrinos via nonrenormalizable operators.

The mixing of up-quarks and charged leptons with the zero-modes of $\psi^c$ 
occurs through bilinear mass terms at the fixed points where \(u,e\)
are contained in  \(\psi^c\vert_{\zeta_\mathrm{PS}} = \mathbf{4}'\), \(\psi^c\vert_{\zeta_\mathrm{GG}}
= {\mathbf{10}^*}'\) and \(\psi^c\vert_{\zeta_\mathrm{fl}} =
(\tilde{\mathbf{5}}',e)\). The mass mixing terms read
\begin{equation}
\begin{split}
W_\mathrm{mix} &= \sum_p m_u^p\ \psi^c\psi \vert_{\zeta_p}  \\
&= m_u^I\ \psi^c\psi \vert_{\zeta_I} + m_u^\mathrm{PS}\ \mathbf{4}' \mathbf{4}^* \vert_{\zeta_\mathrm{PS}} + m_u^\mathrm{GG}\ {\mathbf{10}^*}' \mathbf{10} \vert_{\zeta_\mathrm{GG}} + m_u^\mathrm{fl}\ \tilde{\mathbf{5}}' \tilde{\mathbf{5}}^* \vert_{\zeta_\mathrm{fl}}+ m_e^\mathrm{fl}\ e_{} e^c \vert_{\zeta_\mathrm{fl}}.
\end{split}
\end{equation}
As for the Yukawa couplings, the mixing terms are proportional to the relevant wave function, \(\psi_{++}^{(\alpha)}\) in this case, evaluated at the respective fixed points.
The mass matrices of up-quarks and charged leptons after electroweak
symmetry breaking are then given by
\begin{align}
M_u &= v_1 \sum_{{p = I, \mathrm{GG}}} h_u^p\ \psi^{(i)}_{-+} \psi^{(\alpha)}_{++}\vert_{\zeta_p}\ u_i u^c_\alpha\;  
+ \sum_{{p = I, \mathrm{PS}, \mathrm{GG}, \mathrm{fl}}} \hspace{-1em}
m_u^p\ \psi^{(\alpha)}_{++}\vert_{\zeta_p}\ u_{} u^c_\alpha \notag \\
&\equiv v_1 Y_u^{i\alpha}\ u_i u^c_\alpha\  +\ m_u^\alpha\  u_{} u^c_\alpha\,, \label{MuN}\\
M_e &= v_2 \sum_{p = I, \mathrm{fl}}  h_e^p\ \psi^{(i)}_{--}
\psi^{(\alpha)}_{++}\vert_{\zeta_p}\  e_i e^c_\alpha\
+ \sum_{{p = I, \mathrm{PS}, \mathrm{GG}, \mathrm{fl}}} \hspace{-1em}
m_e^p\ \psi^{(\alpha)}_{++}\vert_{\zeta_p}\ e_{} e^c_\alpha \notag \\ 
&\equiv v_2 Y_e^{i\alpha}\ e_i e^c_\alpha\  +\ m_e^\alpha\ e_{}
e^c_\alpha\,, \label{MeN}
\end{align}
where \(i = 1 \ldots N\), and \(\alpha = 1 \ldots (N+1)\). \(v_{1,2}\)
correspond to the electroweak scale,  while the \(m_u^p\) are assumed
to be of the order of the compactification scale.
At all but the flipped \(SU(5)' \times U(1)'\) fixed points, \(u^c\)
and \(e^c\) are part of the same irreducible representation, forcing
\(m_e^p = m_u^p\) with
\(p = \mathrm{I}, \mathrm{PS}, \mathrm{GG}\); in addition one has the
$SO(10)$ relation \(h_e^\mathrm{I} = h_d^\mathrm{I}\).

The down-quarks \(d_i, d^c_i\) mix with the vectorlike pair \(d, d^c\)
through operators involving the Higgs field \(\Psi \). The superpotential terms read
\begin{equation}
\begin{split}
W_\mathrm{mix} =\ &\lambda_\mathrm{I} \langle \Psi \rangle \psi H_3\vert_{\zeta_I} + \lambda_\mathrm{PS} \langle F^* \rangle \mathbf{4}^* \mathbf{6} \vert_{\zeta_\mathrm{PS}}  \\ 
 &+ \lambda_\mathrm{I}' \langle \Psi^c \rangle \langle H_2 \rangle
 \psi H_4 \vert_{\zeta_I} + \lambda_{\mathrm{PS}}' \langle F \rangle
 \langle \Delta_2 \rangle \mathbf{4}\,
 \mathbf{6}'\vert_{\zeta_\mathrm{PS}} \ ,
\end{split}
\end{equation}
where $H_3\vert_{\zeta_\mathrm{PS}} = \mathbf{6} \supset d$,
$H_4\vert_{\zeta_\mathrm{PS}} = \mathbf{6}' \supset d^c$
and $\Psi\vert_{\zeta_\mathrm{PS}} = F^\ast$. 
All fixed points contribute to a mass term of order the
unification scale for $H_3 H_4 \supset d_{}d^c$. With
\(\langle \Psi \rangle \langle F^\ast \rangle = v_{B-L}\) as \(B-L\) breaking vacuum
expectation value, the down-quark mass matrix becomes
\begin{align}
	M_d &= v_2  h_d^\mathrm{I}\ \psi^{(i)}_{-+}
        \psi^{(j)}_{+-}\vert_{\zeta_\mathrm{I}}\   d_i d^c_j +
          v_{B-L} \ 
\sum_{p = \mathrm{I}, \mathrm{PS}} \lambda_p\ \psi^{(i)}_{+-}\vert_{\zeta_p} \  d_{} d^c_i \\
		   & \qquad {} +  \frac{v_2 v_{B-L}}{M_\ast} \sum_{p =
                     \mathrm{I}, \mathrm{GG}} \lambda_p'\
                   \psi_{-+}^{(i)}\vert_{\zeta_p}\
  d_i d^c + m_D d_{} d^c \notag \\
	    &\equiv v_2 Y_d^{ij}\ d_i d^c_j + v_{B-L} \lambda_d^i\  d_{} d^c_i
            + \frac{v_2 v_{B-L}}{M_\ast} \lambda_d'^i\ d_i d^c + m_D
            d_{} d^c\ , \label{MdN}
\end{align}
where we have explicitly introduced the mass scale \(M_\ast\) for the
nonrenormalizable terms.

A third kind of mixing appears in the neutrino sector. Gauge singlets,
required by the cancellation of gravitational anomalies, can mix with the SM singlet right-handed neutrinos.
These couplings involve the Higgs field \(\Psi^c \).
The gauge singlets can also couple to the left-handed neutrinos
through a nonrenormalizable interaction, involving both \(\Psi \) and
\(H_u\). Considering for simplicity only one singlet \(S\), one
obtains for the bilinear superpotential terms
\begin{equation}
\begin{split}
W_\mathrm{mix} =\ &\kappa_\mathrm{I} \langle \Psi^c\rangle \psi S
\vert_{\zeta_\mathrm{I}} + \kappa_\mathrm{PS} \langle F\rangle\mathbf{4}^* S \vert_{\zeta_\mathrm{PS}} \\
&+ \rho_\mathrm{I}\langle H_1\rangle \langle\Psi\rangle \psi S
\vert_{\zeta_\mathrm{I}}  + \rho_\mathrm{fl} \langle
H_{\tilde{\mathbf{5}}^*}\rangle \langle\tilde T
\rangle \tilde{\mathbf{5}}^* S \vert_{\zeta_\mathrm{PS}} \ ;
\end{split}
\end{equation}
all fixed points contribute to a singlet mass term \(m_S S S\).
Combining the mixing terms with the Yukawa interactions \eqref{eq:scenarioII_Yuk}, one obtains a \(N \times (N+1)\) Dirac neutrino mass matrix,
\begin{align}
M_\nu^D &= v_1 h_{\nu}^\mathrm{I}\ \psi^{(i)}_{--}
\psi^{(j)}_{+-}\vert_{\zeta_\mathrm{I}}\ \nu_i n^c_j + \frac{v_1  v_{B-L}}{M_\ast} 
\sum_{p = \mathrm{I}, \mathrm{fl}} \rho_p \psi^{(i)}_{--}\vert_{\zeta_p}\ \nu_i S\notag \\
		       &\equiv v_1  Y_\nu^{ij} \ \nu_i n^c_j + \frac{v_1
                         v_{B-L}}{M_\ast}  \rho_\nu^i\ \nu_i S\ , \label{MnuN}
\end{align}
where $h_{\nu}^\mathrm{I} = h_{u}^\mathrm{I}$, 
and a \((N+1) \times (N+1)\) Majorana neutrino mass matrix for the
heavy sterile neutrinos,
\begin{align}
	M_n &= \frac{v_{B-L}^2}{M_P} \sum_{p = \mathrm{I}, \mathrm{PS}} \!\!
        h_n^p\ \psi^{(i)}_{+-} \psi^{(j)}_{+-}\vert_{\zeta_p}\ 
  n^c_i n^c_j + v_{B-L}\  \sum_{p = \mathrm{I}, \mathrm{PS}} \!\!
  \kappa_p\ \psi^{(i)}_{+-}\vert_{\zeta_p}\ 
n^c_i S + m_S S S \notag \\
	    &\equiv \frac{v_{B-L}^2}{M_\ast} Y_n^{ij}\  n^c_i n^c_j +
            v_{B-L} \kappa_n^i\ n^c_i S + m_S S S\,. \label{MnN}
\end{align}
Despite the mixing with an additional sterile neutrino, the seesaw
formula can still be applied to obtain the \(N \times N\)  light neutrinos mass matrix
\begin{equation}
	M_\nu = - M_D \frac{1}{M_n} M_D^T\ . \label{seesaw}
\end{equation}

Let us finally emphasize the GUT relations between the parameters. These
are in particular the $SO(10)$ relations for the Yukawa couplings at the
fixed point $\zeta_{\mathrm{I}}$:  $h_{\nu}^\mathrm{I} =
h_{u}^\mathrm{I}$ and $h_{e}^\mathrm{I} = h_{d}^\mathrm{I}$, as well as the relations for the mixing
parameters, \(m_e^p = m_u^p\) with
\(p = \mathrm{I}, \mathrm{PS}, \mathrm{GG}\).

\section{The two heavy families}

Let us now consider the case of two flux quanta, $N=2$, and
apply it to the two heaviest families of the
Standard Model. We start with the down-quark mass matrix, already
discussed in the previous section. As an example, we choose the
parameters\footnote{We also choose $v_{B-L} = \unit[10^{15}]{GeV}$,
$M_* = \unit[2\times 10^{17}]{GeV}$, $\tan\beta = 3$. Note that the
precise values of these parameters is not important. A change can be
compensated by rescaling the superpotential parameters. The parameters
are chosen to reproduce the properties of the two heaviest SM families.}
$h_d^\mathrm{I} = 0.007$, $\lambda_\mathrm{I} = 0.3$,
$\lambda_\mathrm{PS} = 0.2$, $\lambda_\mathrm{I}' = 0.1$, 
$\lambda_\mathrm{GG}' = 0.13$ and $m_D = \unit[10^{15}]{GeV}$. From
Eq.~\eqref{MdN} and the values of the wave functions given in
Eqs.~\eqref{p-+}, \eqref{p+-} one
obtains the $3\times 3$ matrix 
\begin{equation}\label{Md3}
M_d^{(3)} [\mathrm{GeV}] = 
\begin{pmatrix} 0.12\ (1+i) & 0.27 & 0.08 \\ 
0.58\ (1+i) & 1.27 & 0.07 \\ 
7.6\ (1+i)\times 10^{13} & 8.4\times 10^{14} & 1\times
10^{15} \end{pmatrix}\ ,
\end{equation}
where the third row vector contains GUT scale masses.

It turns out that the
quark and charged lepton mass matrices are all of the form
\begin{equation}
\left(\hat{M}^{(N+1)}_{\alpha\beta}\right)  = 
\begin{pmatrix} m_{ij} \quad \mu_i \\ 
M_\beta 
\end{pmatrix}\ , \quad i,j = 1\ldots N\ , \quad \alpha,\beta = 1\ldots
N+1\ ,
\end{equation}
with $m \sim \mu \sim v \ll M \sim v_{B-L}$. This matrix can be reduced to a
$N\times N$ matrix by integrating out one heavy state with GUT-scale
mass $M$ \cite{Asaka:2003iy}. Let us introduce an orthonormal set of
vectors $e_\alpha$ with
\begin{equation}
{e_\alpha}^\dagger e_\beta = \delta_{\alpha,\beta}\, \quad M_\beta = M
(e^\dagger_{N+1})_\beta\ .
\end{equation}
Defining two unitary matrices $V$ and $U$ by
\begin{equation}
\begin{split}
V_{\alpha\beta} &= (e_\beta)_\alpha \ ,\\
U^\dagger_{\alpha\beta} &=\delta_{\alpha,\beta}
-\frac{1}{M}\delta_{\alpha, i}\left(m_{ij}(e_{N+1})_j + \mu_i
  (e_{N+1})_{N+1}\right)\delta_{N+1,\beta} +
\mathcal{O}\left(\frac{1}{M^2}\right)\ ,
\end{split}
\end{equation}
one easily verifies ($i,j = 1\ldots N$),
\begin{equation}\label{intout}
U^\dagger \hat{M} V =
\begin{pmatrix} m_{ik} (e_j)_k + \mu_i (e_j)_{N+1} & 0 \\ 
0 & M
\end{pmatrix}\ , 
\end{equation}
up to corrections of relative order $1/M$. Clearly, the $N\times N$
matrix in the upper-left corner is the relevant low-energy mass matrix.

Using Eq.~\eqref{intout} it is straightforward to integrate out the
heavy state contained in the matrix \eqref{Md3}. For a convenient choice of
vectors $e_i$ one then finds the $2\times 2$ matrix 
\begin{equation}
M_d^{(2)} [\mathrm{GeV}] = 
\begin{pmatrix} 0.12\ (1-i) & 0.15 \\ 
0.57\ (1-i) & 0.87 
\end{pmatrix}
\end{equation}
with the mass eigenvalues
\begin{align}
\hat{M}_d [\mathrm{GeV}]= \mathrm{diag}(0.023, 1.20)\ .
\end{align}
The up-quark mass matrix can be obtained in the same way. We choose the
additional parameters as $h_u^\mathrm{I} = 0.15$, $h_u^\mathrm{GG} =
0.00075$ and $m_u^p =  (0.1,0.05, 0.15, 1.0) \times
\unit[10^{15}]{GeV}$ with $p = (\mathrm{I}, \mathrm{PS}, \mathrm{GG},
\mathrm{fl})$. From Eqs.~\eqref{MuN}, \eqref{p++} and \eqref{p-+} one obtains the $3\times 3$ mass matrix 
\begin{equation}\label{Mu3}
M_u^{(3)} [\mathrm{GeV}] = 
\begin{pmatrix} 11 & 18 & 1.9 \\ 
53 & 81 & 7.1 \\ 
-8.7\times 10^{14} & 4.2\times 10^{14} & 2.0\times
10^{15} \end{pmatrix}\ .
\end{equation}
After integrating out the heavy state one finds the $2\times 2$ matrix 
\begin{equation}
M_u^{(2)} [\mathrm{GeV}] = 
\begin{pmatrix} 11 & 18 \\ 
51 & 83
\end{pmatrix}
\end{equation}
with the mass eigenvalues
\begin{align}
\hat{M}_u  [\mathrm{GeV}]= \mathrm{diag}(0.33, 99)\ .
\end{align}
Diagonalizing the up- and down-quark mass matrices by biunitary
transformations, one obtains from the left-handed rotation matrices
the CKM matrix. We find the mixing angle $\sin\theta^q_{23} = 0.031$.
The obtained bottom, strange, top and charm masses and the CKM mixing
angle agree with the measured values at the GUT scale  within
5\%\footnote{We take all masses and mixings angles at the GUT scale from the recent
  compilation \cite{babu}.}. Given the structure of the quark mass
matrices the small mixing angle $\sin\theta^q_{23}$ is rather
surprizing. Indeed, the individual rotation angles for the
diagonalization of the up- and down-quark matrices are much
larger. However, since the matrices are rather similar, the mismatch
is small, which leads to a small angle in the CKM matrix.

The charged lepton mass matrix is mostly determined by the parameters
of the quark mass matrices. In addition we choose  
$h_e^\mathrm{fl} = 0.0013$ and $m_e^\mathrm{fl} = 1.9 \times
\unit[10^{14}]{GeV}$. From Eq.~\eqref{MeN} and the wave functions given
in Eqs.~\eqref{p++}, \eqref{p--} one obtains the $3\times 3$ charged lepton matrix
\begin{equation}
M_e^{(3)} [\mathrm{GeV}] = 
\begin{pmatrix} 0.22 i & 0.19 + 0.16 i & 0.18 - 0.15 i \\ 
 0.81 - 0.03 i & 1.27 & 0.11 + 0.05 i \\ 
1.1\times 10^{13} & 3.0\times 10^{14} & 5.9\times
10^{14} \end{pmatrix}\ .
\end{equation}
Integrating out the heavy state yields the $2\times 2$ matrix 
\begin{equation}
M_e^{(2)} [\mathrm{GeV}] = 
\begin{pmatrix} 0.23 i & 0.09 + 0.21 i \\ 
0.82 - 0.03 i & 1.07 - 0.02 i
\end{pmatrix}
\end{equation}
with the mass eigenvalues
\begin{align}
\hat{M}_e  [\mathrm{GeV}]= \mathrm{diag}(0.07, 1.38)\ .
\end{align}

The neutrino sector plays a special role. Eq.~\eqref{MnuN} yields a
$2\times 3$ Dirac neutrino mass matrix whereas Eq.~\eqref{MnN} gives
a $3\times 3$ matrix for the heavy sterile neutrinos. Choosing the
remaining parameters as $\rho_\mathrm{I} = 0.8$, $\rho _\mathrm{fl} =
0.1$ and $h_n^\mathrm{I} = 6$, $h_n^\mathrm{PS} = 3$, $m_S = \unit[10^{15}]{GeV}$,
$\kappa_\mathrm{I} =0.3$, $\kappa_\mathrm{PS} = 0.4$, and using the
wave functions
Eqs.~\eqref{p--}, \eqref{p+-}, one finds the Dirac neutrino mass matrix,
\begin{equation}
  M_\nu^D [\mathrm{GeV}] = 
\begin{pmatrix} 
10 i  & 11\ (1+ i) &  0.30 + 0.07 i\\
 37\ (1+i) & 81 &  1.29 + 0.03 i
\end{pmatrix}\ ,
\end{equation}
and the sterile neutrino mass matrix,
\begin{equation}
M_n [\mathrm{GeV}] = 
\begin{pmatrix} 0.53 i & 0.19\ (1 + i) & -0.38\ (1 + i) \\ 
 0.19\ (1 + i) & 1.27 & 5.9 \\ 
-0.38 (1 + i) & 5.9 & 10 \end{pmatrix} \times 10^{14}\ .
\end{equation}
The seesaw formula \eqref{seesaw} then yields the $2\times 2$ matrix
for the light neutrinos
\begin{equation}
M_\nu^{(2)} [\mathrm{meV}] = 
\begin{pmatrix} 2.1 i & 7.6\ (1+i) \\ 
7.6\ (1+i) & 54
\end{pmatrix}
\end{equation}
with the mass eigenvalues
\begin{align}
\hat{M}_\nu  [\mathrm{meV}]= \mathrm{diag}(0, 54)\ .
\end{align}
The vanishing eigenvalue is again a consequence of the dyadic tensor
structure of the neutrino Yukawa matrix.
Diagonalizing the charged lepton and neutrino mass matrices one finds
the MNS mixing angle $\sin\theta^l_{23} = 0.38$. The lepton
masses and the MNS mixing angle are roughly consistent with the
measured values at the GUT scale \cite{babu}. Contrary to the quark sector
the mixing angle is large in the lepton sector. This is a consequence
of the seesaw mechanism which implies that the transitions from flavour
to mass eigenstates  for charged leptons and neutrinos do not compensate
each other.
 
Having fixed the parameters of our model one could hope that changing
the number of flux quanta to $N=3$ would yield a successful
description of the Standard Model with three quark-lepton families. 
Unfortunately, this is not the case. In fact, the third family would
remain massless. The reason is once more the dyadic structure of the
Yukawa matrices. The flux compactification provides a $N$-component
complex vector for each of the six SM fields at
the four fixed points (see Table~2). However, due to the presence of
two Wilson lines, for five SM fields these vectors vanish at two fixed
points. Hence many Yukawa matrices, which are products of two vectors,
and mixings with the additional vector-like states vanish. As a
consequence, mass matrices with full rank can be obtained for $N=2$
but not for $N=3$.

There are several ways to avoid this problem. The simplest possibility
is to have more than one bulk $16$-plet that feels the magnetic flux,
for instance one $16$-plet with charge one and one $16$-plet with
charge two in the case of one flux quantum. Alternatively, one could
start with a smaller bulk gauge group such as $SU(6)$ so that one
Wilson line is enough to achieve symmetry breaking to the Standard
Model. Finally, it is conceivable that not all quarks and leptons
are zero-modes from magnetic flux and that some of them result from
split bulk fields or from fields localized at some fixed points. These
possibilities will be studied in future work.

\section{Supersymmetry breaking and Higgs sector}

For completeness, we summarize in this section some other aspects of
the considered GUT model, which are of phenomenological interest. The
Abelian magnetic flux breaks supersymmetry. Since the matter
hypermultiplet $\psi$ carries $U(1)$ charge, scalar quarks and leptons
acquire universal masses of the order of the compactification scale, which
corresponds to the GUT scale \cite{Bachas:1995ik,Braun:2006se},
\begin{equation}
m^2_{\tilde{q}} = m^2_{\tilde{l}} = \frac{4\pi N}{V_2} \sim (\unit[10^{15}]{GeV})^{-2}\,,
\end{equation}
where $N$ is the number of flux quanta and $V_2$ is the volume of
the compact dimensions. The magnetic flux together with a
nonperturbative superpotential at the fixed points can stabilize the compact dimensions \cite{Braun:2006se},
and it is possible to obtain Minkowski or metastable de Sitter vacua
\cite{Buchmuller:2016dai} with small cosmological constant. In these vacua
the \(U(1)\) vector boson \(A\), the moduli and the axions are all heavy  \cite{Buchmuller:2016bgt},
\begin{equation}
m_\mathrm{moduli} \sim m_\mathrm{axions} \sim m_A\ >\ m_{3/2} \sim
\unit[10^{14}]{GeV}\,.
\end{equation}
The expectation values of the moduli F-terms are of the order of the
gravitino mass. 
Since the moduli dependence of the gauge kinetic terms
is known \cite{Buchmuller:2016bgt}, one easily obtains for the gaugino
masses
\begin{equation}
m_{\tilde{g}} \sim m_{\tilde{w}} \sim m_{\tilde{b}} \sim m_{3/2}\,,
\end{equation}
whereas the SM gauge bosons are massless by construction.

At tree-level the two Higgs fields \(H_u\) and \(H_d\) are massless
since they originate from bulk \(\mathbf{10}\)-plets that do not feel
the magnetic flux. Also the higgsinos are massless since they are
protected by a Peccei-Quinn-type symmetry. One may worry whether it is
possible to consistently extend such a theory up to the GUT scale. In
fact, this is known not to be the case for split supersymmetry or
high-scale supersymmetry because of vacuum instability
\cite{Giudice:2011cg}. On the contrary, for a two-Higgs doublet model,
with or without higgsinos, an extrapolation to the GUT scale is
possible \cite{Lee:2015uza} for stable or
metastable vacua \cite{Bagnaschi:2015pwa}. This
implies constraints on the masses of the heavy neutral scalar and
pseudoscalar as well as charged Higgs bosons \(H\), \(A\),
\(H^\pm\), respectively, and on the ratio of vacuum expectation values,
\begin{equation}
\begin{split}
m_H,\ m_{H^\pm} &= m_A + \mathcal{O}\left(\frac{v^2}{m^2_A}\right)\,,
\quad m_A \gtrsim \unit[1]{TeV}\,, \\ 
\tan\beta &= \frac{v_2}{v_1} = \mathcal{O}(1)\,.
\end{split}
\end{equation}
Solutions of the two-loop renormalization group equations of gauge and
Yukawa couplings show gauge coupling unification at a scale \(\sim
\unit[10^{14}]{GeV}\). At LHC energies the main hope for new discoveries
lies on additional heavy Higgs bosons and higgsinos with masses
\(m_{\tilde{h}} \gtrsim v_{1,2}\). 

A large scale of supersymmetry breaking, like the GUT scale that lies
many orders of magnitude above the electroweak scale, reintroduces the
`hierarchy problem'. Is an enormous fine tuning of parameters needed
to keep the Higgs bosons at the electroweak scale once quantum
corrections are included? This is certainly the case in the
effective 4d theory if only zero-modes are taken into account in the
loop diagrams. However, the situation might be different if the
virtual states are extended to the whole Kaluza-Klein tower. A well
known example is the mass of a Wilson line for an Abelian 6d gauge
field compactified on a torus. After some regularization, one obtains a
Wilson line mass of the order of the inverse size of the compact
dimensions \cite{Antoniadis:2001cv}. Remarkably, as was recently
shown, in the presence of magnetic flux the one-loop quantum corrections to the
Wilson line mass vanish \cite{Buchmuller:2016gib}. This suggests that
magnetic flux may provide a partial protection of scalar masses
w.r.t.\ quantum corrections. However, the applicability to electroweak
symmetry breaking in the considered GUT model remains to be investigated.

\section{Summary and conclusions}
\label{sec:conclusion}
Supersymmetric grand unified theories in higher dimensions provide an
attractive ultraviolet completion of the Standard Model. An example is
the six-dimensional $SO(10)\times U(1)$ model with magnetic flux considered in this
paper, which itself may have an embedding into string theory. An orbifold compactification
to four dimensions with two Wilson lines can break the six-dimensional
gauge symmetry to the Standard Model gauge group. The quark-lepton
families arise as zero-modes of complete $\mathbf{16}$-plets due to the magnetic flux, whereas two
Higgs doublets are obtained as split multiplets from bulk
$\mathbf{10}$-plets. Additional bulk $\mathbf{16}$- and $\mathbf{10}$-plets
lead to further split multiplets that can mix with the complete quark-lepton $\mathbf{16}$-plets.

After describing the symmetry breaking and the quantum numbers of the zero-modes, we have 
analyzed the Yukawa couplings and the bilinear mixings of the
zero-modes. In many flux compactifications, where also the Higgs
fields feel the magnetic flux, Yukawa couplings arise from overlap
integrals of bulk wave functions.  On the contrary, in the model under
consideration Yukawa couplings and mass mixings are superpotential
terms that can arise at the orbifold fixed points. The entire flavour structure is
then contained in complex vectors in flavour space, one for each 
SM representation at each fixed point. These vectors are
a prediction of the flux compactification. They determine the mixings
with split multiplets and their products yield the Yukawa matrices.

We have shown that this pattern of flavour mixing can account for
masses and mixings of the two heaviest Standard Model families. The
mass hierarchies are either due to a hierarchy of Yukawa couplings
or to relative importance of  Yukawa couplings and mass mixings. 
Starting generically from large flavour mixings, the CKM mixing turns
out to be small due to the small mismatch of the rotation matrices for
up- and down-quark mass matrices. On the other hand, the seesaw
mechanism distinguishes between the charged lepton and the neutrino
mass matrices, and the MNS mixing therefore remains large. It turned out
that the two-flavour model cannot be extended to a three-flavour
model in a straightforward way. Due to the dyadic structure of the
Yukawa matrices and the presence of two Wilson lines, too many terms vanish
that would be allowed by the gauge symmetries,  and the
lightest quark-lepton family remains massless. As briefly described in
the previous section, there are 
several ways to avoid this problem. 
The investigation of these possibilities is left for future work.

Let us finally compare the pattern of flavour mixings described in
this paper with successful flavour models of Frogatt Nielsen type
(see, for instance, \cite{Sato:1997hv,Irges:1998ax,Buchmuller:1998zf,
  Altarelli:1998ns}). In these models the entries of the mass matrices
are powers of a small parameter, and therefore small. The powers are
determined by the $U(1)$ charges of the Standard Model
fields. Choosing appropriate charges one obtains hierarchical masses
and small mixings in the quark sector. The lepton sector is different
due to the seesaw mechanism. Mass hierarchies in the Dirac neutrino
mass matrix and the sterile neutrino mass matrix can compensate each other,
leading to small mass ratios and large mixings for the light neutrinos.
On the contrary, in flux compactifications one starts from mass
matrices with large entries, which have rank one due to the dyadic
structure of the Yukawa matrices, and therefore only one massive family. Smaller Yukawa terms and mixing
with split multiplets yields small corrections and increases
the rank. CKM mixings in the quark sector are small due to the small
mismatch between the rotation matrices of up- and down-quark mass
matrices. The neutrino sector is again different because of the seesaw
mechanism and large MNS mixings remain. The two pictures of the origin
of the flavour structure in the Standard Model are complementary to
each other. Mass matrices of Frogatt Nielsen-type 
have been obtained in  heterotic string compactifications (see, for
instance, \cite{bhlr}) as well as F-theory compactifications (see, for
instance, \cite{Heckman:2010bq}). It will be interesting to understand
the connection to string theory compactifications that realize large
flavour mixings as in the flux compactification described in this paper.

\section*{Acknowledgments}
We thank Markus Dierigl and Fabian Ruehle for valuable discussions. This work was supported by the German Science Foundation (DFG) within the Collaborative Research Center (SFB) 676 ``Particles,
Strings and the Early Universe''. 

\newpage

\appendix

\section{Wilson line breaking on orbifolds}

The breaking of $SO(10)$ to the Standard Model gauge group by boundary
conditions is often based on the orbifold $T^2/{(\mathbb Z_2
  \times \mathbb Z_2 \times\mathbb Z_2)}$ \cite{Asaka:2001eh}. For completeness, we
recall the equivalent description based on  
$T^2/{\mathbb Z_2}$ with two Wilson lines in the following.

The torus $T^2$ is obtained by identifying points $y=(y_1,y_2)$ which differ by a
lattice vector, i.e. $T^2 = \mathbb R^2/\mathbb Z^2$
(see Fig.~3),
\begin{equation}
y \sim T_i y = y + \lambda_i\, , \quad i=1,2\ .
\end{equation}
Further identifying points that are related by a rotation of $180^o$
around the origin,
\begin{equation}
y \sim R y = -y\ ,
\end{equation}
one obtains the orbifold $T^2/{\mathbb Z_2}$. The orbifold has
four fixed points,
\begin{equation}
\hat{T}_p \zeta_p = \zeta_p\ ,\quad p= \mathrm{I}, \mathrm{PS}, \mathrm{GG},
\mathrm{fl}\ ,
\end{equation}
where
\begin{equation}
\hat{T}_\mathrm{I} = R\ , \quad \hat{T}_\mathrm{PS} = T_1\circ R\ , \quad
\hat{T}_\mathrm{GG} = T_2\circ R\ , \quad \hat{T}_\mathrm{fl} = T_2\circ T_1\circ R\ .
\end{equation}
In an orbifold field theory one needs an embedding of the space group $S=\{I,R,T_1,T_2\}$ 
into field space (see, for instance, \cite{Hebecker:2001jb}) which is defined on the
covering space $\mathbb R^2$. The corresponding transformations $K[S]$
act linearly in field space. On the fundamental domain of the
orbifold the fields then satisfy certain boundary conditions.

Using the reflection $R$ to break 6d $\mathcal{N}=1$ supersymmetry to
4d $\mathcal{N}=1$ supersymmetry and considering the fields even
under reflection, the embedding of the space group into the gauge
fields is defined by 
\begin{equation}\label{embA}
P_i A(x,\hat{T}_i y)P^{-1}_i = \eta_i A(x,y)\ , \quad i=\mathrm{PS},
\mathrm{GG}\ ,
\end{equation}
where the $P_i$ are $SO(10)$ matrices with
\begin{equation}
P_i^2 = I\ , \quad \eta_i^2 = 1\ .
\end{equation} 
In addition, one has
\begin{equation}
P_\mathrm{fl} A(x,\hat{T}_\mathrm{fl} y)P^{-1}_\mathrm{fl} =
\eta_\mathrm{fl} A(x,y)\ , 
\end{equation}
with $P_\mathrm{fl} =
P_\mathrm{GG}P_\mathrm{PS}$, 
$\eta_\mathrm{fl} = \eta_\mathrm{GG} \eta_\mathrm{PS}$. As the
subscripts indicate, the matrices $P_i$ can be chosen such that  $SO(10)$ is
broken to the Pati-Salam and the Georgi-Glashow subgroups at
$\zeta_\mathrm{PS}$
and $\zeta_\mathrm{GG}$, respectively. 
For hypermultiplets one has
\begin{equation}\label{embphi}
P_i \phi(x,\hat{T}_i y) = \eta_i \phi(x,y)\ , \quad i=\mathrm{PS},
\mathrm{GG}\ ,
\end{equation}
where the matrices $P_i$ depend on the representation of $\phi$, and
 the parities $\eta_i$ can be chosen independently for each hypermultiplet.

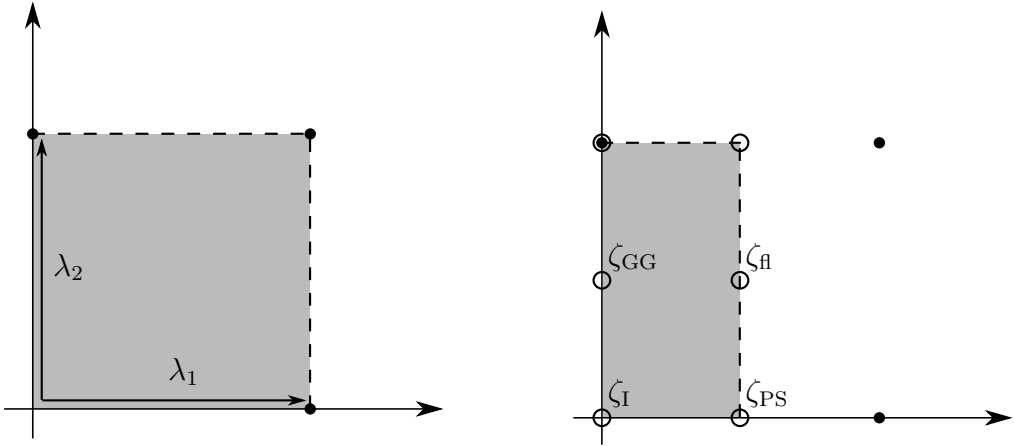
\begin{figure}
\centering 
\begin{overpic}[scale = 1, tics=10]{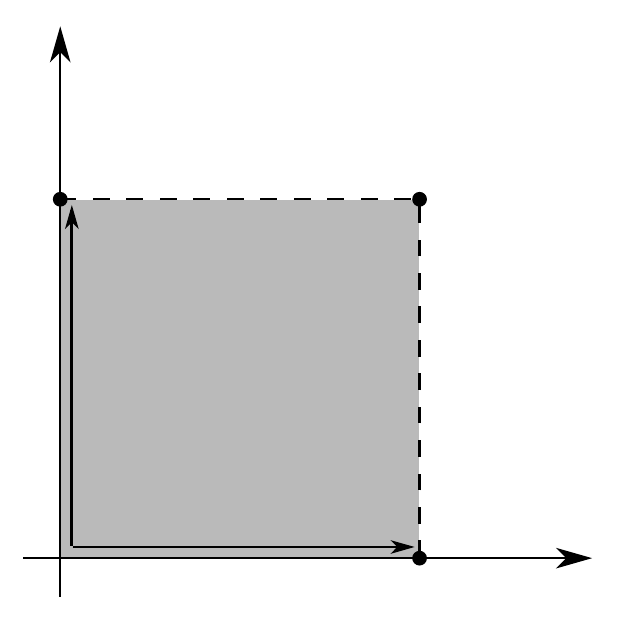}
	\put(38, 16){$\lambda_1$}
	\put(14, 38){$\lambda_2$}
\end{overpic}
\hspace{1cm}
\begin{overpic}[scale = 1, tics=10]{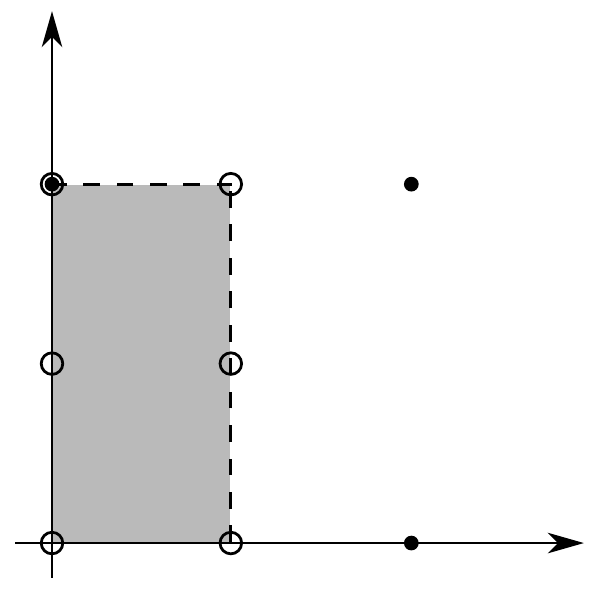}
	\put(10, 12){$\zeta_{\mathrm{I}}$}
	\put(40, 12){$\zeta_{\mathrm{PS}}$}
	\put(40, 42){$\zeta_{\mathrm{fl}}$}
	\put(10, 42){$\zeta_{\mathrm{GG}}$}
\end{overpic}
\caption{Left: torus $T^2$; right: orbifold $T^2/\mathbb Z_2$ with the
fixed points $\zeta_p$, $p= \mathrm{I}, \mathrm{PS}, \mathrm{GG}, \mathrm{fl}$.}
\end{figure}

The embedding \eqref{embA}, \eqref{embphi} of the translations into
field space imply that $SO(10)$ is broken to the
corresponding subgroups at the fixed points,
\begin{equation}
P_i A(x,\zeta_i)P^{-1}_i = \eta_i A(x,\zeta_i) \ ,\quad
P_i \phi(x,\zeta_i y) = \eta_i \phi(x,\zeta_i)\ , 
\quad i=\mathrm{PS}, \mathrm{GG}\ .
\end{equation}
The $SO(10)$ multiplets $A$ and $\phi$ can be decomposed into SM
multiplets, 
$A = \{A^\alpha\}$, $\phi = \{\phi^\beta\}$. Each of them 
belongs to a representation of $G_\mathrm{PS}$ as
well as $G_\mathrm{GG}$ and is therefore characterized by two
parities,
\begin{equation}
A^\alpha (x,\hat{T}_i y) = \eta^\alpha_i A^\alpha (x,y)\ ,
\quad
\phi^\beta (x,\hat{T}_i y) = \eta^\beta_i \phi^\beta(x,y)\ , 
\quad i=\mathrm{PS}, \mathrm{GG}\ .
\end{equation}

\renewcommand{\arraystretch}{1.2}
\begin{table}
\begin{center}
   $\begin{array}[h]{|c||cc|cc|cc|cc|}\hline
 \mbox{SO(10)} &
     \multicolumn{8}{|c|}{ \mathbf{45} }
     \\ \hline
     G_\mathrm{PS} &
     \multicolumn{2}{|c|}{ ( \mathbf{15}, \mathbf{1}, \mathbf{1}) } &
     \multicolumn{2}{|c|}{ ( \mathbf{15}, \mathbf{1}, \mathbf{1}) } &
\multicolumn{2}{|c|}{ ( \mathbf{15}, \mathbf{1}, \mathbf{1}) } &
\multicolumn{2}{|c|}{ ( \mathbf{15}, \mathbf{1}, \mathbf{1}) } 
     \\ \hline
     G_\mathrm{GG} &
     \multicolumn{2}{|c|}{ \mathbf{24}_{0} } &
     \multicolumn{2}{|c|}{ \mathbf{10}_{4} } &
     \multicolumn{2}{|c|}{ \mathbf{10}^\ast_{-4} }  &
     \multicolumn{2}{|c|}{ \mathbf{1}_{0} }
     \\ \hline
 {} &
    \eta_\mathrm{PS} & \eta_\mathrm{GG} &
    \eta_\mathrm{PS} & \eta_\mathrm{GG} &
    \eta_\mathrm{PS} & \eta_\mathrm{GG} &
    \eta_\mathrm{PS} & \eta_\mathrm{GG}
    \\ \hline 
     A &
     + & + &
     + & - &
     + & - &
     + & +
     \\ \hline
   &  \multicolumn{2}{|c|}{G} & \multicolumn{2}{|c|}{} &
     \multicolumn{2}{|c|}{} & \multicolumn{2}{|c|}{X}
     \\ \hline
G_\mathrm{PS} &     
\multicolumn{2}{|c|}{ ( \mathbf{1}, \mathbf{3}, \mathbf{1}) } &
     \multicolumn{2}{|c|}{ ( \mathbf{1}, \mathbf{1}, \mathbf{3}) } &
     \multicolumn{2}{|c|}{ ( \mathbf{1}, \mathbf{1}, \mathbf{3}) }&
\multicolumn{2}{|c|}{ ( \mathbf{1}, \mathbf{1}, \mathbf{3}) }
\\ \hline
 G_\mathrm{GG} &
     \multicolumn{2}{|c|}{ \mathbf{24}_{0} } &
     \multicolumn{2}{|c|}{ \mathbf{24}_{0} } &
     \multicolumn{2}{|c|}{ \mathbf{10}_{4} }  &
     \multicolumn{2}{|c|}{\mathbf{10}^\ast_{-4} }
     \\ \hline
 {} &
    \eta_\mathrm{PS} & \eta_\mathrm{GG} &
    \eta_\mathrm{PS} & \eta_\mathrm{GG} &
    \eta_\mathrm{PS} & \eta_\mathrm{GG} &
    \eta_\mathrm{PS} & \eta_\mathrm{GG}
    \\ \hline 
     A &
     + & + &
     + & + &
     - & + &
     - & +
     \\ \hline
     &  \multicolumn{2}{|c|}{W} & \multicolumn{2}{|c|}{B} &
     \multicolumn{2}{|c|}{} & \multicolumn{2}{|c|}{}
     \\ \hline
G_\mathrm{PS} &     
\multicolumn{2}{|c|}{ ( \mathbf{6}, \mathbf{2}, \mathbf{2}) } &
     \multicolumn{2}{|c|}{ ( \mathbf{6}, \mathbf{2}, \mathbf{2}) } &
     \multicolumn{2}{|c|}{ ( \mathbf{6}, \mathbf{2}, \mathbf{2}) }&
\multicolumn{2}{|c|}{  }
\\ \hline
 G_\mathrm{GG} &
     \multicolumn{2}{|c|}{ \mathbf{24}_{0} } &
     \multicolumn{2}{|c|}{ \mathbf{10}_{4} }  &
     \multicolumn{2}{|c|}{\mathbf{10}^\ast_{-4} }& 
\multicolumn{2}{|c|}{  }
     \\ \hline
 {} &
    \eta_\mathrm{PS} & \eta_\mathrm{GG} &
    \eta_\mathrm{PS} & \eta_\mathrm{GG} &
    \eta_\mathrm{PS} & \eta_\mathrm{GG} &
     & 
    \\ \hline 
     A &
     - & + &
     - & - &
     - & - &
      & 
     \\ \hline
     &  \multicolumn{2}{|c|}{} & \multicolumn{2}{|c|}{} &
     \multicolumn{2}{|c|}{} & \multicolumn{2}{|c|}{}
     \\ \hline\hline
\mbox{SO(10)} &
     \multicolumn{8}{|c|}{ \mathbf{10} }
     \\ \hline
     G_\mathrm{PS} &
     \multicolumn{2}{|c|}{ ( \mathbf{1}, \mathbf{2}, \mathbf{2}) } &
     \multicolumn{2}{|c|}{ ( \mathbf{1}, \mathbf{2}, \mathbf{2}) } &
     \multicolumn{2}{|c|}{ ( \mathbf{6}, \mathbf{1}, \mathbf{1}) } &
     \multicolumn{2}{|c|}{ ( \mathbf{6}, \mathbf{1}, \mathbf{1}) }
     \\ \hline
     G_\mathrm{GG} &
     \multicolumn{2}{|c|}{ \mathbf{5}^\ast{}_{-2} } &
     \multicolumn{2}{|c|}{ \mathbf{5}{}_{+2} } &
     \multicolumn{2}{|c|}{ \mathbf{5}^\ast{}_{-2} }  &
     \multicolumn{2}{|c|}{ \mathbf{5}{}_{+2} }
     \\ \hline
 {} &
    \eta_\mathrm{PS} & \eta_\mathrm{GG} &
    \eta_\mathrm{PS} & \eta_\mathrm{GG} &
    \eta_\mathrm{PS} & \eta_\mathrm{GG} &
    \eta_\mathrm{PS} & \eta_\mathrm{GG}
    \\ \hline 
     H_5 &
     - & + &
     - & - &
     + & + &
     + & -
     \\ 
   &  \multicolumn{2}{|c|}{} & \multicolumn{2}{|c|}{} &
     \multicolumn{2}{|c|}{{D^c}'} & \multicolumn{2}{|c|}{}
     \\ \hline
     H_6 &
     - & - &
     - & + &
     + & - &
     + & +
     \\ 
   &  \multicolumn{2}{|c|}{} & \multicolumn{2}{|c|}{} &
     \multicolumn{2}{|c|}{} & \multicolumn{2}{|c|}{{D}'}
     \\ \hline\hline
    \end{array}$
    \caption{{\rm PS}- and {\rm GG}-parities for the bulk
      $\mathbf{45}$-plet and the
      $\mathbf{10}$-plets $H_5$ and $H_6$. $G$, $W,B$ and $X$ denote
      gluons, electroweak gauge bosons and the additional $U(1)$ gauge
      boson contained in $SO(10)$.}
    \label{tab:P16}
  \end{center}
\end{table}

The matrices $P_\mathrm{PS}$ and $P_\mathrm{GG}$ have been explicitly
given in \cite{Asaka:2001eh} for the fundamental representation. At a fixed point $\zeta_i$ each $SO(10)$
representation can be decomposed into representations of the unbroken
subgroup, and the matrices $P_i$ can be written as linear combinations
of projection operators $\hat{P}$ onto these representations. One finds for
the $\mathbf{45}$-, $\mathbf{10}$- and $\mathbf{16}$-plet, respectively (see  
\cite{Asaka:2002my}),
\begin{align}
P_\mathrm{PS} = & \begin{dcases*}
\hat{P}_{(\mathbf{15},\mathbf{1},\mathbf{1})} +
P_{(\mathbf{1},\mathbf{3},\mathbf{1})} +
P_{(\mathbf{1},\mathbf{1},\mathbf{3})} -
P_{(\mathbf{6},\mathbf{2},\mathbf{2})} 
& for $\mathbf{45}$ \\
 \hat{P}_{(\mathbf{1},\mathbf{2},\mathbf{2})} -
 \hat{P}_{(\mathbf{6},\mathbf{1},\mathbf{1})} 
& for $\mathbf{10}$ \\
 \hat{P}_{(\mathbf{4}^\ast,\mathbf{1},\mathbf{2})} -
\hat{P}_{(\mathbf{4},\mathbf{2},\mathbf{1})}
& for $\mathbf{16}$, \end{dcases*}\\
P_\mathrm{GG} = & \begin{dcases*}
\mathrlap{ \hat{P}_{\mathbf{24}_0} + \hat{P}_{\mathbf{1}_0} - \hat{P}_{\mathbf{10}_4} -
\hat{P}_{\mathbf{10}^\ast_{-4}}}
	\hphantom{
		\hat{P}_{(\mathbf{15},\mathbf{1},\mathbf{1})} +
		P_{(\mathbf{1},\mathbf{3},\mathbf{1})} +
		P_{(\mathbf{1},\mathbf{1},\mathbf{3})} -
		P_{(\mathbf{6},\mathbf{2},\mathbf{2})} }
& for $\mathbf{45}$\\
\hat{P}_{\mathbf{5}^\ast_{-2}} - \hat{P}_{\mathbf{5}_2} & for $\mathbf{10}$\\
\hat{P}_{\mathbf{5}^\ast_3} + \hat{P}_{\mathbf{1}_{-5}} -
\hat{P}_{\mathbf{10}_{-1}} & for $\mathbf{16}$. \end{dcases*} 
\end{align}
Using these relations the parities in Tables~1,2 can be easily determined.

In Table~1 we have listed parities and zero-modes for all fields
that are relevant for the Yukawa couplings and the mixings of split
multiplets with the
$\mathbf{16}$-plets of zero-modes. For completeness we list parities
and zero-modes for the gauge fields and the $\mathbf{10}$-plets
$H_5$ and $H_6$ in Table~3.

\section{Vectors in flavour space}
\label{sec:overlaps}

The flavour structure of our model is entirely determined by the values of the wave functions at the various fixed points,
\(\psi_{a b}^{(i)} \vert_{\zeta_p}\).
The degeneracy index labels idependent fields, which means we can interpret the set of wave functions with given parities as a vector in flavour space,
\begin{equation}
\psi_{a b} = \begin{pmatrix} \psi_{a b}^{(1)} \\ \psi_{a b}^{(2)} \\
  \vdots \end{pmatrix}\ ,\quad a,b = +,-\ .
\end{equation}
In this appendix we give the explicit values of these vectors for \(N = 2\) flux quanta evaluated at the fixed points.

Right-handed up-quarks and electrons are states with even parity at all fixed points.
Therefore, their wave function does not vanish at any fixed point.
The four vectors in flavour space read
\begin{equation}
	\label{p++}
	\begin{alignedat}{4}
	\psi_{++}\vert_{\zeta_\mathrm{I}} &= \begin{pmatrix}0.15  \\ 1.09 \\ 1.68 \end{pmatrix}, &\; \psi_{++}\vert_{\zeta_\mathrm{PS}} &= \begin{pmatrix} 0.15 \\ -1.09 \\ 1.68 \end{pmatrix}, &\;
	\psi_{++}\vert_{\zeta_\mathrm{GG}} &= \begin{pmatrix} 1.68 \\ 1.09 \\ 0.15 \end{pmatrix}, &\; \psi_{++}\vert_{\zeta_{\mathrm{fl}}} &= \begin{pmatrix} 1.68 \\ -1.09 \\ 0.15 \end{pmatrix}.
	\end{alignedat}
\end{equation}
They are three-vectors because the \(\psi_{++}\) are \(N+1\)-fold degenerate.

The left-handed quark doublet's distribution in the internal space is given by \(\psi_{-+}\).
Their wave function is non-zero at two fixed points only,
\begin{equation}
	\label{p-+}
\psi_{-+}\vert_{\zeta_{\mathrm{I}}} = \begin{pmatrix} 0.42 \\ 1.96 \end{pmatrix}, \; \psi_{-+}\vert_{\zeta_{\mathrm{GG}}} = \begin{pmatrix} 1.96 \\ 0.42 \end{pmatrix}.
\end{equation}
The up-quark Yukawa coupling matrix is a linear combination of two
matrices which are obtained as dyadic products of 
flavour space vectors,
\begin{equation}
	Y_u = h_u^I \psi_{++}\vert_{\zeta_\mathrm{I}} \times \psi_{-+} \vert_{\zeta_\mathrm{I}} + h_u^\mathrm{GG} \psi_{++}\vert_{\zeta_\mathrm{GG}} \times \psi_{-+} \vert_{\zeta_\mathrm{GG}}.
\end{equation}
Note that this matrix has rank two, independent of the dimensionality
of the flavour space vectors.
Mixing with the additional state \(u\) is governed by a linear combination of the \(\psi_{++}\vert_{\zeta_p}\), which were given in Eq.~\eqref{p++},
\begin{equation}
m_u = \sum_p m_u^p \psi_{++}^T \vert_{\zeta_p}.
\end{equation}
This gives the overall up-quark mass matrix
\begin{equation}
\mathcal L_m \supset \begin{pmatrix} u^1 & u^2 &  u \end{pmatrix} \begin{pmatrix} v_1 Y_u \\ m_u \end{pmatrix} \begin{pmatrix} u_1^c \\ u_2^c \\ u_3^c \end{pmatrix}.
\end{equation}
The numerical values for an example parameter set are given in Eq.~\eqref{Mu3}.

Right-handed charged leptons have the same field profile as right-handed up quarks.
However, the left-handed lepton fields are given by the wave function \(\psi_{--}\).
They are present at the \(SO(10)\) and flipped \(SU(5)' \times U(1)'\) fixed points,
\begin{equation}
	\label{p--}
\psi_{--} \vert_{\zeta_{\mathrm{I}}} = \begin{pmatrix} 0.28 (1+i) \\ 1.95 \end{pmatrix}, \quad \psi_{--} \vert_{\zeta_{\mathrm{fl}}} = \begin{pmatrix} 0.39 i \\ 1.38 (1 - i) \end{pmatrix}.
\end{equation}
Again, the Yukawa couplings follow from a linear combination of dyadic products,
\begin{equation}
	Y_e = h_d^\mathrm{I} \psi_{++}\vert_{\zeta_\mathrm{I}} \times \psi_{--} \vert_{\zeta_\mathrm{I}} + h_e^\mathrm{fl} \psi_{++}\vert_{\zeta_\mathrm{fl}} \times \psi_{--} \vert_{\zeta_\mathrm{fl}},
\end{equation}
while the mixing with the additional state \( e\) is given by the vectors \(\psi_{++}\vert_{\zeta_p}\),
\begin{equation}
	m_e = \sum_p m_e^p \psi_{++} \vert_{\zeta_p}.
\end{equation}
GUT relations imply \(m_e^p = m_u^p\), \(p = \mathrm{I}, \mathrm{PS}, \mathrm{GG}\).

The down-quark mass terms are
\begin{equation}
\mathcal L_m \supset \begin{pmatrix} d_1 & d_2 &
  d \end{pmatrix} \begin{pmatrix} v_2 Y_d &  \frac{v_2 v_{B-L} }{M_P}
  \lambda_d' \\ v_{B-L} \lambda_d & m_D \end{pmatrix} \begin{pmatrix}
  d^c_1 \\  d^c_2 \\  \hat d^c \end{pmatrix}\ ,
\end{equation}
where \(Y_d = h_d^I \psi_{+-}\vert_{\zeta_I} \times \psi_{-+}\vert_{\zeta_I}  \) is a dyadic product.
The wave function values for right-handed down quarks are
\begin{equation}
	\label{p+-}
\psi_{+-} \vert_{\zeta_\mathrm{I}} = \begin{pmatrix} 0.77 (1 + i) \\
  1.68 \end{pmatrix}, \quad \psi_{+-} \vert_{\zeta_\mathrm{PS}}
= \begin{pmatrix} -0.77 (1+ i) \\ 1.68 \end{pmatrix}\ ,
\end{equation}
while those for the left-handed quarks were already given in Eq.~\eqref{p-+}
The coupling between \( d\) and the \(d^c_j\) is
\begin{equation}
	\lambda_d = \lambda_I \psi_{+-}^T\vert_{\zeta_I} + \lambda_\mathrm{PS} \psi_{+-}^T\vert_{\zeta_\mathrm{PS}},
\end{equation}
while the coupling between the \(d_i\) and \( d^c\) is
\begin{equation}
	\lambda_d' = \lambda_I' \left. \psi_{-+} \right\vert_{\zeta_I}
        + \lambda_{GG}' \left. \psi_{-+}
        \right\vert_{\zeta_{\mathrm{GG}}}\ .
\end{equation}

The Dirac neutrino mass matric reads
\begin{equation}
\mathcal L_m \supset \begin{pmatrix} \nu_1 & \nu_2 \end{pmatrix} \begin{pmatrix} v_1 Y_\nu & \frac{v_1 v_{B-L}}{M_P} \kappa_\nu \end{pmatrix} \begin{pmatrix} \nu^c_1 \\ \nu^c_2 \\ S \end{pmatrix},
\end{equation}
with the Yukawa matrix \(Y_\nu = h_u^I \psi_{--}\vert_{\zeta_I} \times \psi_{+-}\vert_{\zeta_I}\) given by the dyadic product of the vectors given in Eq.~\eqref{p--} and Eq.~\eqref{p+-}.
The mixing with the \(SO(10)\) singlet scalar \(S\),
\begin{equation}
	\rho_\nu = \rho_I \left. \psi_{--} \right\vert_{\zeta_I} + \rho_\mathrm{fl} \left. \psi_{--} \right\vert_{\zeta_\mathrm{fl}}
\end{equation}
is also governed by the values in Eq.~\eqref{p--}.

Finally, the Majorana mass matrix for the sterile neutrinos is
\begin{equation}
\mathcal L_m \supset \begin{pmatrix} \nu^c_1 & \nu^c_2 &
  S \end{pmatrix} \begin{pmatrix} \frac{v_{B-L}^2}{M_*} M_{\nu^c} &
  \frac{v_{B-L}}{2} \kappa_{\nu} \\ \frac{v_{B-L}}{2} \kappa_{\nu}^T &
  m_N \end{pmatrix} \begin{pmatrix} \nu^c_1 \\ \nu^c_2 \\
  S \end{pmatrix}\ .
\end{equation}
The corresponding mass matrix \(M_{n}\) reads 
\begin{equation}
	M_{n} = h_n^I \psi_{+-}\vert_{\zeta_I} \times \psi_{+-}\vert_{\zeta_I} + h_n^\mathrm{PS} \psi_{+-}\vert_{\zeta_\mathrm{PS}} \times \psi_{+-}\vert_{\zeta_\mathrm{PS}},
\end{equation}
where the \(\psi_{+-}\vert_{\zeta_p}\) are already defined in Eq.~\eqref{p+-}.
The same flavour space vectors govern the mixing with the sterile neutrino \(S\) 
\begin{equation}
	\kappa_{\nu} = \kappa_I \left. \psi_{+-} \right\vert_{\zeta_I} + \kappa_\mathrm{PS} \left. \psi_{+-} \right\vert_{\zeta_\mathrm{PS}}.
\end{equation}

This explicit presentation showcases that the entire flavour sector is determined by the four sets of wave functions.
These determine flavour space vectors when evaluated at the fixed points.
All relevant numerical values were presented in Eqs.~\eqref{p++}, \eqref{p-+}, \eqref{p--} and \eqref{p+-}.

\providecommand{\href}[2]{#2}\begingroup\raggedright\endgroup

\end{document}